\documentclass[final]{siamltex}
\usepackage{amssymb,amsfonts,amsmath,epsfig,pstricks}
\newcommand{\tstep}{\Delta t}
\newcommand{\enz}{{\mathbb N}_0}

\newcommand{\non}{\nonumber}

\newcommand{\pdhfrac}[2]{\mathchoice{\frac{#1}{#2}}{#1/#2}{#1/#2}{#1/#2}}

\newcommand{\beqa}{\begin{eqnarray}}
\newcommand{\eeqa}{\end{eqnarray}}
\newcommand{\beqas}{\begin{eqnarray*}}
\newcommand{\eeqas}{\end{eqnarray*}}
\newcommand{\beq}{\begin{equation}}
\newcommand{\eeq}{\end{equation}}

\title{Analysis of the two-regime method on square meshes}

\author{Mark B. Flegg\thanks{Mathematical Institute, University of Oxford,
24-29 St. Giles', Oxford, OX1 3LB, United Kingdom; 
e-mails: flegg@maths.ox.ac.uk, chapman@maths.ox.ac.uk, 
erban@maths.ox.ac.uk}
 \and \quad S. Jonathan Chapman$^*$
 \and \quad Likun Zheng\thanks{Department of Mathematics,
      University of California, Irvine, 540-N, Rowland Hall,
      Irvine, CA 92697, USA; e-mail: likunz@uci.edu}
 \and \qquad\qquad Radek Erban$^*$}
\begin{document}

\maketitle

\begin{abstract}
The two-regime method (TRM) has been recently developed for optimizing 
stochastic reaction-diffusion simulations \cite{Flegg:2012:TRM}. It is 
a multiscale (hybrid) algorithm which uses stochastic reaction-diffusion
models with different levels of detail in different parts of the 
computational domain. The coupling condition on the interface between 
different modelling regimes of the TRM was previously derived for 
one-dimensional models. In this paper, the TRM is generalized 
to higher dimensional reaction-diffusion systems. Coupling 
Brownian dynamics models with compartment-based models on regular (square) 
two-dimensional lattices is studied in detail. 
In this case, the interface between different 
modelling regimes contain either flat parts or right-angled corners. 
Both cases are studied in the paper. For flat interfaces, it is shown 
that the one-dimensional theory can be used along the line perpendicular 
to the TRM interface. In the direction tangential to the interface,
two choices of the TRM parameters are presented. Their applicability
depends on the compartment size and the time step used in
the molecular-based regime. The two-dimensional generalization 
of the TRM is also discussed in the case of corners.
\end{abstract}



\pagestyle{myheadings}
\thispagestyle{plain}

\section{Introduction}
There are two common approaches to stochastic reaction-diffusion modelling:
(i) compartment-based models; and (ii) molecular-based models 
\cite{Erban:2007:PGS,Erban:2009:SMR}. Molecular-based models provide a higher
level of detail, but they are often more computationally intensive
than compartment-based models. In some applications, microscopic detail 
is only required in a relatively small region, for example, close to the 
cellular membrane or a particular organelle 
\cite{Flegg:2012:DSN,Erban:2012:MRS}.
Such problems are best simulated by a hybrid method which uses a detailed 
modelling approach in localized regions of particular interest 
(in which accuracy and microscopic detail is important) and a less 
detailed model in other regions in which accuracy may be traded 
for simulation efficiency. To apply this general idea to stochastic
reaction-diffusion modelling, one has to introduce a suitable boundary
condition between different modelling regimes. In \cite{Flegg:2012:TRM}, 
we derived the appropriate boundary condition for coupling 
one-dimensional compartment-based and molecular-based models. We
developed the two-regime method (TRM) which has the 
accuracy of the detailed molecular-based approach (in the region where 
it is required), but benefits from the efficiency of a less detailed (coarser) 
compartment-based model in other parts of the computational domain. 
In this paper, we generalize the TRM to higher dimensional
simulations.

In the remainder of this first section, we introduce the notation 
which is used during the rest of this manuscript. We summarize both 
compartment-based (Section \ref{seccomp}) and molecular-based modelling
(Section \ref{secmole}). Then we introduce the TRM in Section \ref{sectrm}. 
Our main results are presented and derived in Section \ref{2Dresults}. 
In Section \ref{examples}, we demonstrate the applicability of the
presented theory using several illustrative numerical examples.
Finally, we discuss other multiscale (hybrid) stochastic reaction-diffusion
approaches from the literature. These methods vary in implementation 
and applicability and we put them into context against the TRM
in Section \ref{secdiscussion}.

\subsection{Compartment-based modelling}\label{seccomp} Mesoscale 
compartment-based modelling of reaction-diffusion processes begins 
by partitioning the domain $\Omega_C$ into compartments (open sets) 
$\mathcal{C}_j$, $j=1,\ldots,K,$ such that the compartments do not 
overlap and they cover the whole domain 
$\Omega_C$ (i.e. $\cup_{j=1}^K \overline{\mathcal{C}}_j = \Omega_C$ 
and $\mathcal{C}_i \cap \mathcal{C}_j = \emptyset $, for $i\neq j$,
where overbars denote the closure of the corresponding set).  
Assuming that there are $M$ chemical species $\mathcal{Z}_i$, 
$i=1,\ldots,M$, the state of the system is completely defined 
by the copy numbers $\mathcal{N}_{i,j} \in \enz$ of molecules for chemical 
species $\mathcal{Z}_i$ found in the compartment $\mathcal{C}_j$, 
$i=1,\ldots,M$, $j=1,\ldots,K$. In what follows, symbol
$\enz$ denotes the set of nonnegative integers, i.e.
$\enz \equiv \{0,1,2,3,\dots\}.$ The simulation of reaction and 
diffusion of the molecules in the system is usually implemented 
by event-driven algorithms, which include the Gillespie algorithm 
\cite{Gillespie:1977:ESS}, the Next Reaction Method \cite{Gibson:2000:EES}
or the Next Subvolume Method \cite{Hattne:2005:SRD}. They 
have been implemented in several open-source software packages 
including MesoRD \cite{Hattne:2005:SRD}, URDME \cite{Engblom:2009:SSR},
STEPS \cite{Wils:2009:SMS} and SmartCell \cite{Ander:2004:SFS}.
In this paper, we will use a derivative of the Next Reaction Method 
from Gibson and Bruck \cite{Gibson:2000:EES}. 

Event-driven algorithms require the calculation of event propensities 
for a particular system state \cite{Gillespie:1977:ESS}. An event 
propensity $\alpha_{\mathcal{E}}$ 
is the rate (per unit time) for an event $\mathcal{E}$ to occur that 
changes the state of the system. Events in compartment-based 
reaction-diffusion processes may include: reaction events (in which 
chemical molecules of some species change into molecules of other 
species, or are just introduced or removed from the system), diffusive 
events (in which molecules of a chemical species jump from one 
compartment to an adjacent compartment) or boundary events (in which 
molecules are absorbed by, react with or reflect from a domain boundary). 
A putative time $t_\mathcal{E}$ for each event $\mathcal{E}$ can be 
found given some current time $t$ using
\begin{equation}\label{update}
 t_\mathcal{E} = t + \frac{1}{\alpha_\mathcal{E}}
 \ln\left( \frac{1}{r_\mathcal{E}} \right),
\end{equation}
where $r_\mathcal{E}$ are uniformly distributed random 
numbers between 0 and 1 chosen separately for each occurence of each event. 
The next event which takes place in the system is determined 
by finding which event corresponds to time 
$\min_\mathcal{E} t_\mathcal{E}$ where the minimum 
is taken over the set of all possible events \cite{Gibson:2000:EES}. 
The state is 
changed to reflect the occurence of the event and the current 
time is then updated to $t: = \min_\mathcal{E} t_\mathcal{E}$.
The current event might also change propensities of some 
related events. The putative times for these events must 
therefore be scaled to reflect the change in propensity.
That is, 
\begin{equation}\label{puttimetc}
t_\mathcal{E}^\mathrm{new} := 
t + \frac{\alpha_\mathcal{E}^\mathrm{old}}{\alpha_\mathcal{E}^\mathrm{new}}
\left(
t_\mathcal{E}^\mathrm{old} - t
\right), 
\end{equation}
where $\alpha_\mathcal{E}^\mathrm{old}$ and 
$\alpha_\mathcal{E}^\mathrm{new}$ 
(resp. $t_\mathcal{E}^\mathrm{old}$ and $t_\mathcal{E}^\mathrm{new}$)
are the propensities (resp. putative times)
for event $\mathcal{E}$ before and after the current event
takes place \cite{Gibson:2000:EES}. A putative time for the next 
occurence of the current event must be resampled using 
formula (\ref{update}). The simulation is then constructed 
by a series of successive events over time, in each instance, 
defined by the most imminent event and performing the state change 
that defines that event.

\subsubsection{Reaction events}
One of the main assumptions of compartment modelling of 
stochastic reaction-diffusion models is that each compartment 
is small enough that it may be considered well mixed \cite{Erban:2007:PGS}. 
Reactions are modelled in each compartment by defining 
the propensity for reaction in each compartment. Consider 
the reaction $\mathcal{R}$ in compartment $\mathcal{C}_j$ 
given by the general form 
\begin{equation}
\sum_{i=1}^M \beta_i \mathcal{Z}_i \xrightarrow{\kappa} 
\sum_{i=1}^M \gamma_i \mathcal{Z}_i,
\label{genreact}
\end{equation}
where $\beta_i \in \enz$ (resp. $\gamma_i \in \enz$) are the 
numbers of molecules of chemical $\mathcal{Z}_i$, $i=1,2,\dots,M$, 
that are required as reactants (resp. products) of the chemical reaction 
$\mathcal{R}$ and $\kappa$ is the reaction rate. 
We define the notation for this event $\mathcal{E}=(\mathcal{R},j)$. 
In realizing this reaction event in the $j$-th compartment, 
the number of molecules $\mathcal{N}_{i,j}$, $i=1,2,\dots, M,$ 
$j = 1,2,\dots,K$, change by the corresponding stoichiometric 
coefficient $\nu_i = \gamma_i - \beta_i$. Considering mass action 
chemical kinetics, the propensity for this event to occur depends 
on the number of available reactants in the 
compartment $\mathcal{C}_j$. In 3D, we can postulate this dependence 
in the following form \cite{Erban:2007:PGS,Erban:2009:SMR}
\begin{equation}
 \alpha_{\mathcal{R},j} 
 = \kappa V_j^{1 - \sum_i \beta_i}
\prod_{i=1}^M \frac{\mathcal{N}_{i,j}!}{(\mathcal{N}_{i,j}-\beta_i)!}
\label{genprop}
\end{equation}
where $V_j$ is the volume of the compartment $\mathcal{C}_j$. 
Table \ref{tab1} shows the examples of propensity 
$\alpha_{\mathcal{R},j}$ for some simple reactions 
$\mathcal{R}$ in compartment $\mathcal{C}_j$ and the 
subsequent changes to the state of the system 
that occur as a result of the reaction event. 

\begin{table}
\caption{Examples of propensities and effects of reactions 
in compartments. To simplify this table, we omit species 
with zero coefficients in reaction and product complexes which
were included in the general form $(\ref{genreact})$.
}\label{tab1}
\centering
\begin{tabular}{|c|c|c|}
 \hline
\raise -2mm \hbox{\rule{0pt}{6mm}} example reaction $\mathcal{R}$ 
  & $\alpha_{\mathcal{R},j}$ 
  & changes of the state vector \\
\hline \hline
\raise -7mm \hbox{\rule{0pt}{16mm}} $\mathcal{Z}_i+\mathcal{Z}_k \xrightarrow{\kappa} \mathcal{Z}_l$ 
& $\displaystyle \frac{\kappa\mathcal{N}_{i,j}\mathcal{N}_{k,j}}{V_j}$ 
& \begin{tabular}{c}$\mathcal{N}_{i,j}$ changes to $\;\mathcal{N}_{i,j}-1$ 
\\  
$\mathcal{N}_{k,j}$ changes to $\;\mathcal{N}_{k,j}-1$ \\ 
$\mathcal{N}_{l,j}$ changes to $\;\mathcal{N}_{l,j}+1$ 
\end{tabular}\\
\hline
\raise -4mm \hbox{\rule{0pt}{10mm}}
$2\mathcal{Z}_i\xrightarrow{\kappa} \mathcal{Z}_l$ 
&  $\displaystyle \frac{\kappa\mathcal{N}_{i,j}(\mathcal{N}_{i,j}-1)}{V_j}$ 
& \begin{tabular}{c}$\mathcal{N}_{i,j}$ changes to 
$\;\mathcal{N}_{i,j}-2$ \\ 
$\mathcal{N}_{l,j}$ changes to $\;\mathcal{N}_{l,j}+1$ 
\end{tabular}\\
\hline
\raise -7mm \hbox{\rule{0pt}{16mm}} 
$\mathcal{Z}_i\xrightarrow{\kappa} \mathcal{Z}_k+\mathcal{Z}_l$ 
& $\kappa\mathcal{N}_{i,j}$ & 
\begin{tabular}{c}$\mathcal{N}_{i,j}$ changes to 
$\;\mathcal{N}_{i,j}-1$ \\  
$\mathcal{N}_{k,j}$ changes to $\;\mathcal{N}_{k,j}+1$ 
\\ $\mathcal{N}_{l,j}$ changes to $\;\mathcal{N}_{l,j}+1$ 
\end{tabular}\\
\hline
\raise -2mm \hbox{\rule{0pt}{7mm}} 
$\emptyset \xrightarrow{\kappa} \mathcal{Z}_l$ 
& $\kappa V_j$ &  $\mathcal{N}_{l,j}$ changes to $\;\mathcal{N}_{l,j}+1$ \\
\hline
\end{tabular}
\end{table}

The main focus of this paper is on 2D simulations. In this case, the 
formula (\ref{genprop}) has to be slightly modified by replacing the
compartment volume $V_j$ by its area. Since the propensity
$\alpha_{\mathcal{R},j}$ is dimensionless, some rate constants have 
different physical units in 2D than in 3D. In applications, 2D simulations 
in the domain $\Omega_C \subset {\mathbb R}^2$ are often viewed as a model 
of a real 3D domain $\Omega_C \times (0,w)$ where the domain width $w$ 
is so small that the spatial distribution along the third axis can 
be neglected. Then the formula (\ref{genprop}) can be applied with 
the standard interpretation (i.e. physical units) of the reaction rates. 
The area of each 2D compartment is multiplied by $w$ to get the corresponding  
volume $V_j$ in (\ref{genprop}).
 
\subsubsection{Diffusion events}

Diffusion in compartment-based models of reaction-diffusion processes 
is defined by the stochastic jumping of molecules of chemical species 
$\mathcal{Z}_i$, $i=1,2,\dots,M$, between any two adjacent compartments 
from $\mathcal{C}_j$ to $\mathcal{C}_k$,
$j,k=1,2,\dots,K$. We shall define the notation for this diffusive 
event to be $\mathcal{E}=(\mathcal{D},i,j,k)$. The propensity for 
a diffusive event is given by
\begin{equation}
\label{diffprop}
 \alpha_{\mathcal{D},i,j,k} = q_{j,k} D_i \, \mathcal{N}_{i,j},
\end{equation}
where $q_{j,k}$ is dependent on the morphology and relative 
positions of the compartments $\mathcal{C}_j$ and $\mathcal{C}_k$,
and $D_i$ is the diffusion constant for chemical $\mathcal{Z}_i$. 
For adjacent square or cubic compartments $\mathcal{C}_j$ and 
$\mathcal{C}_k$ of length $h$ on a regular lattice, we have
$q_{j,k} = 1/h^2$, i.e.
\begin{equation}
\label{q}
\alpha_{\mathcal{D},i,j,k} = \frac{D_i}{h^2} \, \mathcal{N}_{i,j},
\end{equation}
and $q_{j,k} = 0\;$ if $\;\mathcal{C}_j$ and 
$\mathcal{C}_k$ do not share a common side. 
For more irregular compartment shapes, $q_{j,k}$ can be 
determined by the finite element discretisation of the 
diffusion equation on a lattice whose vertices are at 
the centers of the compartments \cite{Engblom:2009:SSR}. 
During the diffusive event $(\mathcal{D},i,j,k)$, 
the state of the system is changed to reflect the movement 
of one molecule of $\mathcal{Z}_i$ from $\mathcal{C}_j$ 
to $\mathcal{C}_k$, i.e.
$\mathcal{N}_{i,j}$ changes to
$\mathcal{N}_{i,j} - 1$ and 
$\mathcal{N}_{i,k}$
changes to $\mathcal{N}_{i,k} + 1$.  

\subsubsection{Boundary events}
Boundary events are caused by the diffusion of mole\-cules 
into the boundaries of the domain $\partial \Omega_C$. 
It is no surprise then that boundary events are linked 
closely to diffusion events. Consider the compartment 
$\mathcal{C}_j$ which is adjacent to $\partial \Omega_C$. 
A diffusive event of a molecule of chemical $\mathcal{Z}_i$ 
which would ordinarily result in a jump from $\mathcal{C}_j$ 
to a compartment on the other side of $\partial \Omega_C$ 
actually results in an interaction of the molecule with the 
boundary. Such an interaction usually results in one of 
two outcomes either: the molecule is absorbed; or the 
molecule is reflected \cite{Erban:2007:RBC}. Molecules
attached to the surface can be released back into the solution
\cite{Andrews:2005:SRL,Lipkova:2011:ABD}. However, 
in this paper, we will consider reflective boundary conditions on
all external boundaries for simplicity.

\subsection{Molecular-based modelling}
\label{secmole}
Molecular-based approaches to reaction-diffusion modelling
are characterized by the prescription 
of exact coordinates in space for each molecule of each 
chemical species $\mathcal{Z}_i$, $i = 1,2,\dots, M$, in 
the continuous domain $\Omega_M \subset {\mathbb R}^N$, where
$N=1,2,3$. The trajectory of large molecules (such as proteins) 
are computed using Brownian dynamics 
\cite{Andrews:2004:SSC,Franz:2012:MRA,vanGunsteren:1982:ABD}.
A random displacement of each molecule every 
timestep $\tstep$ models the effect of solvent molecules 
on the large molecules of interest without the need to 
simulate each solvent molecule individually. We will denote 
the $j$-th molecule of chemical species $\mathcal{Z}_i$ as 
$Z_i^j$. Given a time step $\tstep$ the position 
${\mathbf x}_{i,j}(t+\Delta t)$ of molecule $Z_i^j$
at time  $t + \Delta t$ is computed from its position 
${\mathbf x}_{i,j}(t)$ at time $t$ in the $N$-dimensional 
continous space $\Omega_M$ by
\begin{equation}\label{updateeqn}
\mathbf{x}_{i,j}(t + \Delta t) 
= 
\mathbf{x}_{i,j}(t) + \sqrt{2D_i\tstep} \, {\boldsymbol{\zeta}},
\end{equation} 
where ${\boldsymbol{\zeta}} \in \mathbb{R}^N$ 
is a vector of $N$ uncorrelated normally distributed random 
numbers with zero mean and unit variance chosen separately for 
each molecule. The equation (\ref{updateeqn}) is a discretization 
of the standard Brownian motion
\begin{equation}
\label{standardBD}
\mbox{d} \mathbf{X}_{i,j}
= 
\sqrt{2D_i} \; \mbox{d}\mathbf{W_{i,j}}.
\end{equation}
Molecular-based models have been implemented in several software
packages, including Smoldyn \cite{Andrews:2004:SSC}, MCell
\cite{Stiles:2001:MCM} and GFRD \cite{vanZon:2005:GFR}. They have
been used for modelling several biological systems, including the signal
transduction in bacterium {\it E. coli} \cite{Lipkow:2005:SDP} 
and the MAPK pathway \cite{Takahashi:2010:STC}.

\subsubsection{Chemical reactions} It is relatively straigtforward
to implement zero-order and first-order chemical reactions
in molecular-based models \cite{Erban:2007:PGS}.
There is a variety of different ways to model bimolecular
(second-order) molecular-based reactions. Consider two molecules $Z_i^j$ and 
$Z_k^l$ which can react according to the following bimolecular
reaction
$$
\mathcal{Z}_i+\mathcal{Z}_k \xrightarrow{\kappa} \mathcal{Z}_m.
$$
Then a suitable probability of reaction per time step  
is chosen such that the macroscopic rate of reaction between 
the two chemicals is $\kappa$ \cite{Erban:2009:SMR}.
This probability is a function of the distance 
$|\mathbf{x}_{i,j}-\mathbf{x}_{k,l}|$ between reacting
molecules. Models postulate that the 
molecules can only react if they are within a specific
distance (reaction radius) \cite{Andrews:2004:SSC}. 
Care must also be taken when generating the initial positions 
of products of chemical reactions. This is especially the 
case whenever reversible reactions are present in the system. 
If the molecules are not initialized properly then the products 
may unphysically react immediately after being created 
\cite{Lipkova:2011:ABD,Khokhlova:2012:GFR}. 

\subsubsection{Boundary conditions}
\label{secbcmodel}
Molecules that migrate over domain boundaries are treated 
depending on whether they are reflective, absorbing or reactive 
boundaries \cite{Erban:2007:RBC}. When modelling boundary conditions,
one has to take into account that (\ref{updateeqn}) is only
an approximation of the Brownian motion (\ref{standardBD}).
Let us consider that the formula (\ref{updateeqn}) gives the 
updated position of the molecule $Z_i^j$ inside the domain
$\Omega_M$, i.e. ${\mathbf x}_{i,j}(t+\Delta t) \in \Omega_M$.
Then there is still a nonzero probability that the molecule
(which follows (\ref{standardBD})) left the domain $\Omega_M$ 
during the time $(t,t+\tstep)$ and then returned back to 
$\Omega_M$. We denote this probability by 
$P_m \equiv P_m({\mathbf x}_{i,j}(t),{\mathbf x}_{i,j}(t+\tstep))$.
Near a flat boundary $\partial \Omega_M$, probability
$P_m$ takes the analytical form
\begin{equation}\label{Pm}
P_m 
= 
\exp\left( 
\frac{- \mbox{dist} ({\mathbf x}_{i,j}(t), \partial \Omega_M) \;
        \mbox{dist} ({\mathbf x}_{i,j}(t+\tstep), \partial \Omega_M)}
	{D_i \tstep}
\right),
\end{equation}
where $\mbox{dist}(\cdot,\partial \Omega_M)$ represents the 
distance from the boundary $\partial \Omega_M$ \cite{Andrews:2004:SSC}.
Whilst boundaries are sometimes 
only implemented at time $t + \Delta t$ for molecule $Z_i^j$ 
if ${\mathbf x}_{i,j}(t+\Delta t)$ computed by (\ref{updateeqn})
is outside $\Omega_M$, thorough implementation of boundary 
conditions should not only be considered when 
${\mathbf x}_{i,j}(t+\Delta t) \notin \Omega_M$ but 
also with the probability $P_m$ if 
${\mathbf x}_{i,j}(t+\Delta t) \in \Omega_M$. 
No closed form solution exists for $P_m$ for irregular boundary 
geometries $\partial \Omega_M$. In practice, a curved boundary is 
usually described locally by a flat approximation which 
increasingly becomes more accurate for small values of $\tstep$
\cite{Andrews:2004:SSC}. 

\subsection{The two-regime method}
\label{sectrm}

Molecular-based techniques are usually prefered over compartment-based 
simulation techniques when the concentration of molecu\-les is low to give 
a level of microscopic detail that is not achieveable in the mesoscopic 
compartment-based approaches. However, it can be very cumbersome 
numerically to simulate every single molecule and perform probability 
tests for every pair of molecules that have the potential to react if 
the copy numbers of molecular species are large. 
In such cases, compartment-based approaches 
are appropriate. They provide a level of efficiency that does not require 
the tracking of each individual molecule. This comes, however, at the 
cost of detail in the simulation. This paper focuses on the TRM which 
provides a way to spatially connect regions that use compartment-based 
modelling in regions that require a mesoscopic level of detail with 
that of molecular-based modelling in regions where the concentrations 
are lower or a high level of detail is required \cite{Flegg:2012:TRM}.

The TRM for simulation of stochastic reaction-diffusion processes 
is characterized by its partition of the computational domain 
$\Omega \subset {\mathbb R}^N$, $N = 1, 2, 3$, into two non-overlapping 
open subsets $\Omega_C$ and $\Omega_M$, i.e. 
$\Omega_C \cap \Omega_M = \emptyset$ and
$\overline{\Omega}_C \cup \overline{\Omega}_M = \overline{\Omega}$ 
where overbars denote the closure of the corresponding set.
We denote by $I$ the interface between the subdomains
$\Omega_C$ and $\Omega_M$, i.e.
$I = \partial\Omega_C \cap \partial\Omega_M$. 
Internally, molecules are simulated in subdomains
$\Omega_C$ and $\Omega_M$ by compartment-based 
and molecular-based approaches which were described 
in Sections \ref{seccomp} and \ref{secmole},
respectively. Molecules in $\Omega_M$ are updated 
at prescribed times separated by $\tstep$. Meanwhile, 
molecules in $\Omega_C$ are updated at the events 
determined by compartment modelling rules described 
in Section \ref{seccomp}. The simulation is thus 
built from a series of time steps that occur at each 
``regular time step" separated by $\tstep$ and each event 
inside $\Omega_C$. As the time updates can be classified by 
the region that they apply to, updates corresponding to 
compartment-based events are known as $C$-events (or compartment events) 
and the regular updates separated by $\tstep$ in time are known 
as $M$-events. There are several possible variants
of the TRM \cite{Flegg:2012:TRM}. In this paper, we will analyse
the TRM in the form which is summarized in Table \ref{TRMalg}.
\begin{table}
\caption{The pseudocode of the TRM for stochastic 
reaction-diffusion simulation.}
\label{TRMalg}
\begin{center}
$\boxed{\hbox{\vbox{\hsize=0.95\hsize
\parindent -7.5mm
\leftskip 7.5mm
\rightskip 0.2mm
{\bf (i)} \hskip 2.3mm 
Define the subdomains $\Omega_C$ and $\Omega_M$ and 
the interface $I = \partial \Omega_C \cap \partial \Omega_M$. Subdivide 
$\Omega_C$ into compartments $\mathcal{C}_j$, $j=1,\ldots,K$.
Choose the time step $\tstep$ between updates of the 
molecular-based regime ($M$-events) in $\Omega_M$. 
\hfill\hfill\hfill\par 
{\bf (ii)} \hskip 1mm 
Specify the initial condition in $\Omega_M$ by 
placing molecules $Z_i^j$ in $\Omega_M$ 
at initial positions ${\mathbf x}_{i,j}(0)\in\Omega_M$,
$i=1,2,\dots,M$, $j = 1,2,\dots,n(i)$, where
$n(i)$ is the initial number of molecules of the $i$-th
chemical species $\mathcal{Z}_i$ in $\Omega_M$.
\hfill\hfill\hfill\par 
{\bf (iii)} 
Specify the initial condition in $\Omega_C$ by 
initializing the copy numbers $\mathcal{N}_{i,j}$ in $\Omega_C$ 
for each chemical species $\mathcal{Z}_i$ in each compartment 
$\mathcal{C}_j$, $i=1,2,\dots,M$, $j = 1,2,\dots,K.$
Initialize time as $t:=0$. 
\hfill\hfill\hfill\par 
{\bf (iv)} \hskip 0.3mm 
Use equation (\ref{update}) to calculate $t_\mathcal{E}$, 
the putative times at which all $C$-events $\mathcal{E}$
will take place. Set $t_M = \Delta t$ and 
$t_C = \min_\mathcal{E} t_\mathcal{E}$ where the minimum 
is taken over all possible $C$-events $\mathcal{E}$.
\hfill\hfill\hfill\par
{\bf (v)} \hskip 1.1mm 
If $t_C \le t_M$, then the next $C$-event occurs: 
\hfill\hfill\hfill\par 
\parindent -5mm
\leftskip 15.5mm
$\bullet$ \; Update current time $t := t_C$. \hfill\hfill\hfill\par 
$\bullet$ \; Change the number of molecules in $\Omega_C$ to reflect 
the specific $C$-event that has occured. If this event is 
one in which a molecule of chemical species $\mathcal{Z}_i$ 
leaves $\Omega_C$ bound for $\Omega_M$, 
then compute its initial position in $\Omega_M$ according 
to the probability distribution $f_{i,j}(\mathbf{x})$ and
remove it from the corresponding compartment $\mathcal{C}_j$.
\hfill\hfill\hfill\par 
$\bullet$ \; 
Calculate the next putative time for the current $C$-event 
by equation, (\ref{update}). For all propensity functions 
$\alpha_\mathcal{E}$ that are changed as a result of the $C$-event, 
determine the putative times of the corresponding event by 
equation (\ref{puttimetc}). \hfill\hfill\hfill\par 
$\bullet$ \; Set $t_C := \min_\mathcal{E}(\tau_\mathcal{E})$.
\hfill\hfill\par
\parindent -7.5mm
\leftskip 7.5mm
{\bf (vi)} \hskip 0.6mm
If $t_M \le t_C$, then the next $M$-event occurs: 
\hfill\hfill\hfill\par 
\parindent -5mm
\leftskip 15.5mm
$\bullet$ \; Update current time $t : = t_M$. 
\hfill\hfill\hfill\par 
$\bullet$ \; Change the locations of all molecules in $\Omega_M$ using 
equation (\ref{updateeqn}). 
\hfill\hfill\hfill\par 
$\bullet$ \; 
Implement boundary conditions
at the external boundary $\partial \Omega_M \setminus I$.
\hfill\hfill\hfill\par 
$\bullet$ \; Initialize all molecules which migrated from $\Omega_C$
to $\Omega_M$ since the previous $M$-event at locations 
computed in the step (v) according to $f_{i,j}(\mathbf{x})$.
\hfill\hfill\hfill\par 
$\bullet$ \; Perform all reaction events in $\Omega_M$. 
\hfill\hfill\hfill\par 
$\bullet$ \; Identify all molecules that interact with the interface 
$I$ from $\Omega_M$ (excluding those just initiated) using 
conditions (a)--(b) from Section \ref{secomc}.
Move each molecule into the appropriate compartment in 
$\Omega_C$ with probability $\Psi$. Otherwise, its position is
reflected back to $\Omega_M$. 
\hfill\hfill\hfill\par 
$\bullet$ \; For all propensity functions $\alpha_\mathcal{E}$ that are 
changed as a result of the $M$-event 
determine the putative times of the corresponding $C$-event by equation
 (\ref{puttimetc}). \hfill\hfill\hfill\par 
$\bullet$ \; Update $t_M := t_M + \tstep$ and, if necessary, set 
$t_C := \min_\mathcal{E}(\tau_\mathcal{E})$. 
\hfill\hfill\hfill\par
\parindent -7.5mm
\leftskip 7.5mm
{\bf (vii)} \hskip 0.1mm
Repeat steps (v) and (vii) until the desired end of the simulation.
\hfill\hfill\hfill\par }}}$
\end{center}
\end{table}
In the step {\bf (i)}, we define $\Omega_C$, $\Omega_M$ and the
time step $\Delta t$. Initial conditions in $\Omega_M$ and
$\Omega_C$ are implemented in the steps {\bf (ii)} and {\bf (iii)},
respectively. In the step {\bf (iv)}, we also define putative times
$t_M$ and $t_C$ when the next $M$-event and $C$-event will occur,
respectively. Then the TRM repeats steps {\bf (v)} and
{\bf (vi)} until the desired end of the simulation.

\subsubsection{Transition of molecules from $\Omega_C$ to $\Omega_M$}
The partition of the domain into $\Omega_C$ and $\Omega_M$ is an 
artificial partitioning that should not interfere with the natural 
diffusion of molecules in the domain. To describe migration
of molecules from $\Omega_C$ to $\Omega_M$, we need an expression for 
the propensity of molecules to jump from compartments $\mathcal{C}_j$, 
adjacent to the interface $I$, into $\Omega_M$. When a molecule 
successfully jumps from $\mathcal{C}_j$ into $\Omega_M$, it 
must be placed with a specific set of coordinates by virtue 
of the modelling approach in $\Omega_M$. For a regular array of 
square or cubic compartments in $\Omega_C$, we define the propensity  
of chemical species $\mathcal{Z}_i$ to jump from $\mathcal{C}_j$ 
into $\Omega_M$ to be some multiple $\Phi_{i,j}$ times the natural 
jumping propensity between neighbouring compartments 
(provided that compartment $\mathcal{C}_j$ is adjacent to $I$, a so-called 
{\it interfacial compartment}, otherwise this propensity is equal to $0$). 
That is (compare with (\ref{q})),
\begin{equation}\label{qbar}
\alpha_{\mathcal{D},i,j,M} = \Phi_{i,j} \frac{D_i}{h^2} \, \mathcal{N}_{i,j},
\end{equation}
where the index $j$ is a reference to the originating interfacial 
compartment $\mathcal{C}_j$. Subsequently, the molecule, after 
being chosen to jump from $\mathcal{C}_j$ into $\Omega_M$ is 
initialized at a position $\mathbf{x}\in\Omega_M$. We do not 
restrict this initialization to a specific location but rather 
consider that the initial position is chosen from a probability 
distribution $f_{i,j}(\mathbf{x})$. 

\subsubsection{Transition of molecules from $\Omega_M$ to $\Omega_C$}
\label{secomc}
In step {\bf (vi)}, a molecule originating in $\Omega_M$ is transfered 
into $\Omega_C$ with a probability $\Psi \in [0,1]$ if the molecule 
interacted with the interface $I$ within the time interval $[t,t+\tstep]$. 
It is 
postulated that molecule $Z_i^j$ interacted with $I$ if one of these
two conditions is satisfied:

\leftskip 1.5cm
\medskip

{\parindent -6mm
(a) ${\mathbf x}_{i,j}(t+\tstep)$ computed by (\ref{updateeqn})
satisfies ${\mathbf x}_{i,j}(t+\tstep)\in\Omega_C$;

(b) ${\mathbf x}_{i,j}(t) \in \Omega_M$ and $r \le P_m$ where
    $r$ is uniformly distributed random number in $(0,1)$
    and $P_m \equiv 
    P_m({\mathbf x}_{i,j}(t),{\mathbf x}_{i,j}(t+\tstep))$
    was introduced in Section \ref{secbcmodel}.
    For a straight interface $I$, the probability $P_m$ is given 
    by (\ref{Pm}).
\par}
        
\leftskip 0cm   
\medskip

\noindent
If the probability $\Psi$ is strictly less than 1, then we have to 
incorporate into the TRM that all molecules which satisfy (a) and which
are not transported to $\Omega_C$ are reflected back to $\Omega_M$.
However, this condition is not necessary in 1D where it is possible
to prove that $\Psi=1$ \cite{Flegg:2012:TRM}. This simplifies the 
implementation of the TRM in 1D. In the next section, we summarize 
the results of the 1D theory presented in \cite{Flegg:2012:TRM}.
Then, in Section \ref{2Dresults}, we analyse the TRM in 2D.
 
\subsubsection{Summary of analysis of the TRM in 1D}
In \cite{Flegg:2012:TRM}, the TRM is presented and analysed in 
a one dimensional 
domain $\Omega = (-\infty,\infty)$ which was divided
by the interface $I = \{0\}$ into $\Omega_C = (-\infty,0)$
and $\Omega_M = (0,\infty)$. The subdomain $\Omega_C$
was divided into compartments of the same length $h$, i.e.
$\mathcal{C}_j = (-jh,(1-j)h)$, $j=1,2,\ldots$. 
In this case, there is only one interfacial compartment corresponding 
to $j=1$ and coordinates $x\in\Omega_M$ are defined as the 
displacement from the interface $I$. It was obtained that
\begin{equation}
 \Phi_{i,1} = \frac{2h}{\sqrt{\pi D_i \tstep }},
\qquad \qquad \qquad \Psi = 1,
\label{results1d}
\end{equation}
\begin{equation}
\label{trans1d} 
f_{i,1}(x) = 
\sqrt{\frac{\pi}{4D_i\tstep}}\, \mathrm{erfc}
\left( \frac{x}{\sqrt{4D_i\tstep}} \right), \qquad \ x\in\Omega_M,
\end{equation}
where $\mathrm{erfc}(x) = 2/\sqrt{\pi} \int_x^\infty \exp(-t^2) \mbox{d}t$ 
is the complementary error function. These formulae were derived
under the assumption $D\tstep \sim h^2$ as $h \rightarrow 0$, a condition
which is used throughout the manuscript. 
If $\tstep$ gets larger than this then the expected jump distance of
molecules in the Brownian domain, $\sqrt{2D\tstep}$, becomes greater than the
compartment size $h$. Since in the simplest form of the TRM scheme we
do not allow Brownian particles to jump to interior compartments
(particles that cross the interface are placed in a boundary
compartment), we often assume $D\tstep < h^2$ (we are usually
interested in a finer resolution in the Brownian domain than the
compartment domain).
If $D \tstep$ is chosen much smaller than $h^2$ then the coupling 
between domains is still accurate in one dimension.
However, we will see that choosing  $\sqrt{2D\tstep}
\ll h$ in higher dimensions can create some 
numerical artefacts associated with molecules
diffusing along the interface $I$.            

\section{Main results} 
\label{2Dresults}
In the rest of this paper we wish to derive forms for $\Phi_{i,j}$, 
$\Psi$ and $f_{i,j}(\mathbf{x})$ in domains with dimensions greater 
than $N=1$. Since these parameters, the sole requirements for the 
correct spatial coupling of regions $\Omega_C$ and $\Omega_M$, 
are independent of the parameters for reactions we do not need to 
note the chemical species in our analysis. We shall therefore drop 
the index $i$ and relabel $D_i$ as $D$ which will denote the diffusion 
constant for the chemical species in question. We therefore wish to find the 
parameters $\Phi_{j}$, $\Psi$ and $f_{j}(\mathbf{x})$ for domains 
that have more than one dimension. We shall limit the analysis to 
regularly spaced square compartments of length $h$ for dimension
$N=2$, but the results can be easily generalized to regularly spaced 
$N$-dimensional cubic compartments (see Section \ref{Ndim}). 
The parameters will be derived for a flat interface $I$ and then 
we will discuss the case where $I$ may have a corner.  
 
\subsection{Matching at a flat interface in 2D}\label{flat2d}
\begin{figure}[t]
\centerline{\epsfig{file=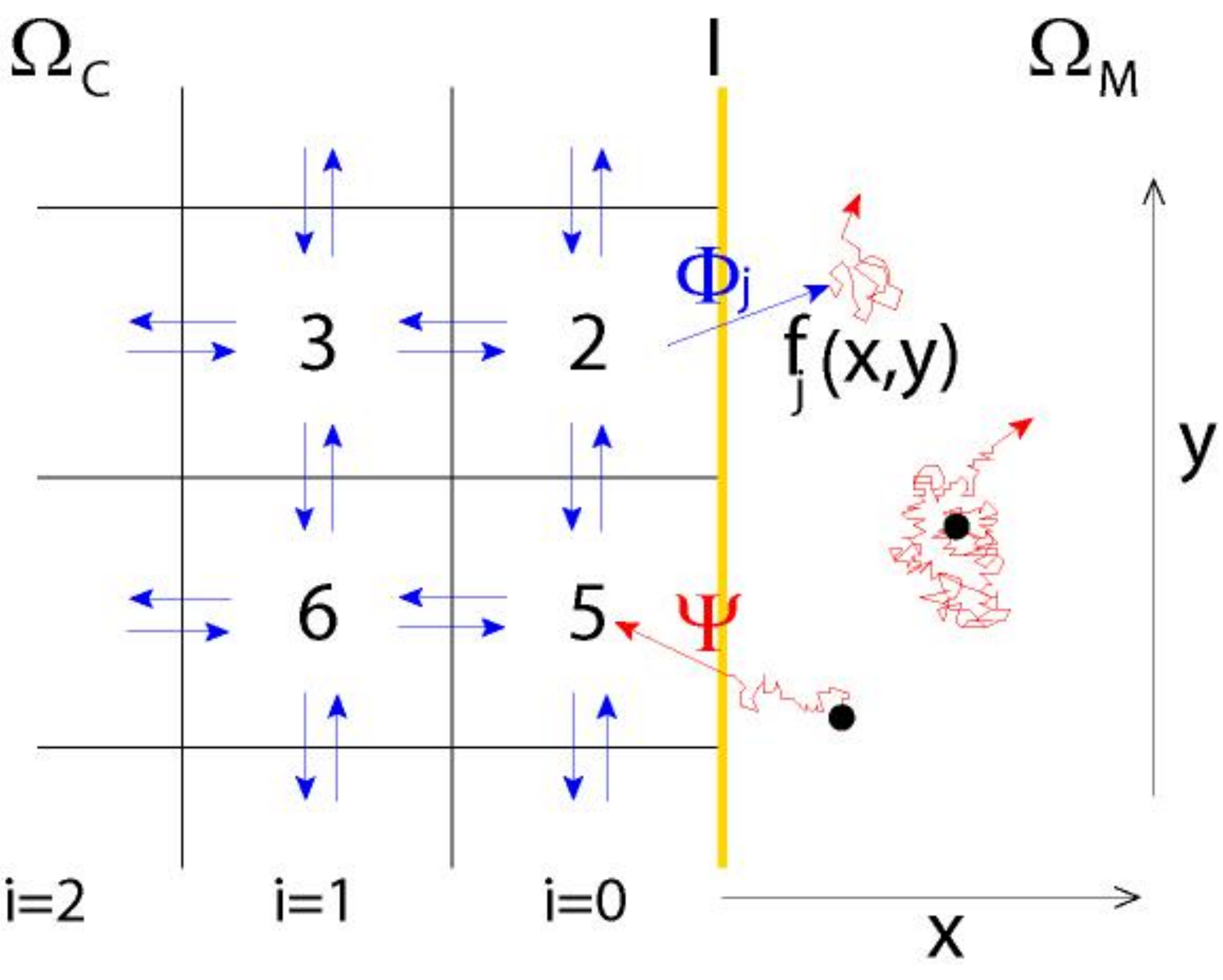,height=5cm}}
\caption{Graphical representation of the TRM in 2D for a flat interface $I$.
Compartment-based regime $\Omega_C$ is on the left (numbers denote the
number of molecules in the corresponding compartment),
molecular-based regime $\Omega_M$ on the right (three illustrative
trejectories of individual molecules are plotted as red lines).
The interface $I$ is plotted as a yellow line.}
\label{diagram2D}
\end{figure}

The considered geometry is represented graphically in Figure \ref{diagram2D}.
We present here a derivation to 
the parameters $\Phi_{j}$, $\Psi$ and $f_{j}(\mathbf{x})$ for the 
TRM applied to an infinite two-dimensional 
domain $\Omega = \mathbb{R}^2$ and a 
single flat interface $I$. To derive these algorithm parameters
(and formulate them in a reasonably simplified form),
we denote by $(x,y)$ the Cartesian coordinates 
that describe domain $\Omega$. Without loss of generality we assign 
$\Omega_M$ to the region defined by $x > 0$ and therefore $\Omega_C$ 
to the region defined by $x < 0$ (i.e. the interface $I$ is the line 
$x=0$). The 
compartments $\mathcal{C}_j$ are regularly spaced squares of side 
length $h$. We find it convenient to describe the compartments 
by two indicies such that each 
compartment $\mathcal{C}_{i,j}$ is assigned to the region described 
by $-(i+1)h < x < -ih$ and $jh-h/2 < y < jh+h/2$, where 
$i \in \enz$ and $j \in {\mathbb Z}$. In what follows, we will
denote by $p_{i,j}(t) h^2$ the probability of a molecule to be in
the compartment $\mathcal{C}_{i,j}$, i.e. 
$p_{i,j}(t)$ is the (discretized, averaged)
probability density in the compartment $\mathcal{C}_{i,j}$. 

Since the analysis of the TRM only depends on the properties of diffusion
\cite{Flegg:2012:TRM}, we can write the governing equations 
as the evolution equations for the probability density of a single 
diffusing molecule in $\Omega$. The goal of the TRM is to correctly
approximate its probability density function $P(x,y,t)$ where
$(x,y) \in \Omega$ and time $t \ge 0$. Using equations 
(\ref{q}), (\ref{updateeqn}), (\ref{Pm}), (\ref{qbar}) and 
the description of the evolution of molecules near the interface 
$I$ during the TRM, the TRM master equations for both the average probability 
density $p_{0,n}(t)$ for a molecule to be in the $\mathcal{C}_{0,n}$ 
compartment and the probability density $p(x,y,t)$ in $\Omega_M$ 
are given by
\begin{eqnarray}
\nonumber 
p_{0,n}(t+\tstep) & = & 
\left( 1- \frac{(3+\Phi_0)D\tstep}{h^2}\right)p_{0,n}(t) 
+ 
\frac{D\tstep}{h^2} \big( p_{0,n-1}(t) + p_{0,n+1}(t) + p_{1,n}(t) \big) 
\\ 
& + & \frac{2\Psi}{h^2} \!
\int_{nh-h/2}^{nh+h/2} \!\!\!\mathrm{d}y \int_{-\infty}^{\infty} 
\!\!\!\mathrm{d}\bar{y} \int_{0}^{\infty} \!\!\!\mathrm{d}x \int_{0}^{\infty} 
\!\!\!\mathrm{d}\bar{x} \ p(\bar{x},\bar{y},t) \, K\left( x+\bar{x}
\right)K\left( y-\bar{y}\right), \qquad
\label{blah1}
\end{eqnarray}
\begin{eqnarray}
p(x,y,t+\tstep) &=&  
\int_{-\infty}^{\infty} \!\!\mathrm{d}\bar{y} \int_{0}^{\infty} 
\!\!\mathrm{d}\bar{x} \ p(\bar{x},\bar{y},t)\left[K\left( x-\bar{x}\right) 
+ (1-2\Psi)K\left( x+\bar{x}\right)\right]   K\left( y-\bar{y}\right) 
\nonumber \\
& + &
D \tstep \sum_{j \in {\mathbb Z}} \Phi_j f_j(x,y) p_{0,j}(t) ,
\label{blah2}
\end{eqnarray}
where $K(x) = (4\pi D\tstep)^{-1/2}\exp(-x^2/(4D\tstep))$ is 
the distribution of the random displacement 
in the position of molecules between $t$ and $t+\tstep$ given
by (\ref{updateeqn}). One key assumption that is made in equation 
(\ref{blah1}) is that molecules that are absorbed into $\Omega_C$ from 
$\Omega_M$ are absorbed into the closest compartment $\mathcal{C}_{0,j}$ 
to their position calculated at $t+\Delta t$ (see the limits of 
integration over the variable $y$ in (\ref{blah1})). It can also 
be shown \cite{Erban:2007:RBC,Flegg:2012:TRM}
that the contributions of the two ways a molecule may be 
absorbed by the interfacial compartment (see cases (a)--(b)
in Section \ref{sectrm}) to the total number of absorbed molecules 
in a time step are equal giving rise to the factor of $2$ in the 
integral term on the right hand side of equation (\ref{blah1}) and 
the absorption component of the integral on the right hand side of 
equation (\ref{blah2}).  

We may further simplify equations (\ref{blah1})--(\ref{blah2})
by using the symmetry of the domain in the $y$-direction. 
Firstly, since $\Phi_{j}$ (where now $j$ refers to compartments 
$\mathcal{C}_{0,j}$ on the interface) is dependent only on the morphology 
of $\mathcal{C}_{0,j}$ and its relative position to $I$, we expect 
$\Phi_{j}$ to be independent of $j$ and we therefore denote 
\begin{equation}
\Phi_{j} \equiv \Phi.
\label{symsimplicity}
\end{equation}
Secondly, symmetry in the $y$-direction 
also allows us to make the conclusion that 
\begin{equation}
f_j(x,y) = f_0(x,y - j h).
\label{secondsymsimplicity}
\end{equation}
where the index of $f_j$ refers only to interfacial 
compartments $\mathcal{C}_{0,j}$. 

In the vicinity of $x=0$ there 
is a boundary layer of width $O(\sqrt{\tstep})$
\cite{Erban:2007:RBC,Flegg:2012:TRM}. We rescale equations 
(\ref{blah1})--(\ref{blah2}) 
using the dimensionless boundary layer coordinate $x=\xi\sqrt{D\tstep}$. 
We also denote the probability density and 
placement function in this boundary layer region by 
$p_{\mathrm{in}}(\xi,y,t)$ and 
$f_{\mathrm{in},j}(\xi,y) = \sqrt{D\tstep}\, f_j(\xi\sqrt{D\tstep},y)$.
The rescaling of $f_j$ is necessary to satisfy the 
normalisation condition 
$$
\int_{-\infty}^{\infty} \int_{0}^{\infty} f_j(x,y) \ \mathrm{d}x 
\ \mathrm{d}y = 1
$$ since (as we will see) $f_j$ vanishes outside 
of the boundary layer. Thus, using (\ref{symsimplicity})
and (\ref{secondsymsimplicity}),
in the boundary layer coordinates 
equations (\ref{blah1})--(\ref{blah2}) become
\begin{eqnarray}
\nonumber 
p_{0,n}(t+\tstep) & = & 
\left( 1- \frac{(3+\Phi)}{\Lambda^2}\right)p_{0,n}(t) 
+ 
\frac{1}{\Lambda^2} \big( p_{0,n-1}(t) + p_{0,n+1}(t) + p_{1,n}(t) \big) 
\\ 
& + & \frac{2\Psi}{h \Lambda} \!
\int_{nh -h/2}^{nh+h/2}
\!\!\!\mathrm{d}y \int_{-\infty}^{\infty} 
\!\!\!\mathrm{d}\bar{y} \int_{0}^{\infty} \!\!\!\mathrm{d}\xi 
\int_{0}^{\infty} 
\!\!\!\mathrm{d}\bar{\xi} \ p_{\mathrm{in}}(\bar{\xi},\bar{y},t) \, 
\kappa\left( \xi+\bar{\xi}\right)K\left( y-\bar{y}\right), 
\qquad
\label{blah3}
\end{eqnarray}
\begin{eqnarray}
p_{\mathrm{in}}(\xi,y,t+\tstep) &=&  
\int_{-\infty}^{\infty} \!\!\mathrm{d}\bar{y} \int_{0}^{\infty} 
\!\!\mathrm{d}\bar{\xi} \ p_{\mathrm{in}}(\bar{\xi},\bar{y},t)
\left[\kappa\left( \xi-\bar{\xi}\right) 
+ (1-2\Psi)\kappa\left( \xi+\bar{\xi}\right)\right]   
K\left( y-\bar{y}\right) 
\nonumber \\
& + &
\sqrt{D\tstep} \,\Phi \sum_{j \in {\mathbb Z}} f_{\mathrm{in},0}(\xi,y-jh) 
\, p_{0,j}(t) ,
\label{blah4}
\end{eqnarray}
where $\Lambda = \pdhfrac{h}{\sqrt{D\tstep}}$ and 
$\kappa(\xi) = \sqrt{D\tstep}\, K(\sqrt{D\tstep} \, \xi)
= (4\pi)^{-1/2} \exp(-\xi^2/4)$.
Let us denote $\bar{p}(y,t) = P(0,y,t)$ and $\bar{p}_x(y,t) = P_x(0,y,t)$
where $P(x,y,t)$ is the distribution
which the TRM approximates (for $x \in {\mathbb R}$ and $y \in 
{\mathbb R}$).
In order for the models to join smoothly at the interface $I$ 
we require on the compartment-based side that 

\begin{equation}\label{blah5}
\begin{split} 
 p_{0,n}(t) & \sim P(0,nh,t) = \bar{p}(nh,t), \\
 p_{1,n}(t) & \sim P(-h,nh,t) = \bar{p}(nh,t) - h\bar{p}_x(nh,t) + O(h^2), \\
 p_{0,n+1}(t) & \sim P(0,nh+h,t) = 
 \bar{p}(nh+h,t), \\
 p_{0,n-1}(t) & \sim P(0,nh-h,t) 
 = \bar{p}(nh-h,t), \\
\end{split}
\end{equation}
while, for the molecular-based side, we require no rapid variation 
in the boundary layer, so that
\begin{equation}\label{bleh5}
\begin{split} 
 p_{\mathrm{in}}(\xi,y,t) &  \sim \bar{p}(y,t) 
 + \sqrt{D\tstep}(\xi+C_x)\bar{p}_x(y,t) 
  + \ldots, \\
  p_{\mathrm{in}}(\xi,y,t+\tstep) &  \sim \bar{p}(y,t+\tstep) 
 + \sqrt{D\tstep}(\xi+C_x)\bar{p}_x(y,t) 
  + \ldots,
  \end{split}
\end{equation}
where we have allowed for a small shift $C_x$ in which the molecular-based 
region ``sees'' the interface \cite{Flegg:2012:TRM}.
Similarly
\[\int_{-\infty}^\infty K\left( y-\bar{y}\right)
\bar{p}(\bar{y},t)\,\mathrm{d}\bar{y}  = \bar{p}(y,t) + O(\tstep). \]
Substituting the expansions (\ref{blah5})--(\ref{bleh5})
into equations (\ref{blah3}) and (\ref{blah4}), using $h \sim
\sqrt{D\tstep}$ as  $h \to 0$ and $\tstep \to 0$, we obtain,   
\begin{eqnarray}
\nonumber 
0 & = & 
- \frac{\Phi}{\Lambda^2}\bar{p}(nh,t)
 + 
\frac{1}{\Lambda^2}  h\bar{p}_x(nh,t)
\\ 
& + & \frac{2\Psi}{ \Lambda}\frac{1}{\sqrt{\pi}} 
\bar{p}(nh,t)
 +\frac{2\Psi}{ \Lambda}\sqrt{D\tstep}\, \left(\frac{1}{2} +
   \frac{C_x}{\sqrt{\pi}}\right)\bar{p}_x(nh,t) + O(h^2),
\qquad
\label{blah3.3}\\
 0 &=&  - \Psi \, \mathrm{erfc}\left(
    \frac{\xi}{2}\right)
\bar{p}(y,t)\non \\
&& \mbox{ }
 +\sqrt{D\tstep}\left(
\frac{(2-2 \Psi)}{\sqrt{\pi}}e^{-\xi^2/4}  - C_x \Psi
\mathrm{erfc}\left( \frac{\xi}{2}\right)   + (\Psi-1) \xi
\mathrm{erfc}\left( \frac{\xi}{2}\right) 
\right)
\bar{p}_x(y,t)   
\nonumber \\
& & \mbox{ }+ 
\sqrt{D\tstep} \,\Phi \sum_{j \in {\mathbb Z}} f_{\mathrm{in},0}(\xi,y-jh) 
\, \bar{p}(jh,t) + O(h^2).
\label{blah4.2}
\end{eqnarray}
Equating coefficients of $\bar{p}(nh,t)$ and $\bar{p}_x(nh,t)$ in
(\ref{blah3.3}) gives
\begin{equation}\label{ratiophis}
\Phi =  \frac{2\Psi \Lambda}{\sqrt{\pi}},\qquad
 1 = \Psi\left(1 + \frac{2C_x}{\sqrt{\pi}} \right).
\end{equation}
From the coefficient of $\bar{p}_x$ in  (\ref{blah4.2}) we see that
\begin{equation}
\Psi = 1, \quad \mbox{and} \quad C_x = 0.
\label{psicxvalue}
\end{equation}
Consequently, equation (\ref{ratiophis}) implies that
$\Phi$ satisfies (\ref{results1d}). Using (\ref{psicxvalue}) and
(\ref{blah4.2}) we find that $f_{\mathrm{in},0}$ 
has to satisfy
\begin{eqnarray*}
  \mathrm{erfc}\left(
    \frac{\xi}{2}\right)
\bar{p}(y,t) &=&  
 \frac{2 h}{\sqrt{\pi}} \sum_{j \in {\mathbb Z}} f_{\mathrm{in},0}(\xi,y-jh)
\, \bar{p}(jh,t)+O(h^2).
\end{eqnarray*}
Writing 
\[  f_{\mathrm{in},0}(\xi,y) = \frac{\sqrt{\pi}}{2} \mathrm{erfc}\left(
    \frac{\xi}{2}\right) F(y),\] 
this becomes 
\begin{eqnarray}
 \bar{p}(y,t) &=&  
h \sum_{j \in {\mathbb Z}} F(y-jh) 
\, \bar{p}(jh,t) +O(h^2).
\label{blah4.5}
\end{eqnarray}
This is a standard interpolation problem. The simplest weight function which
gives $O(h^2)$ accuracy is the triangle function 
\begin{equation}\label{simple} F(y) = \left\{ \begin{array}{ll}
\displaystyle \pdhfrac{1}{h}\left(1-\pdhfrac{|y|}{h}\right), & 
\mbox{for} \; -h < y < h,\\[3mm]
0, & \mbox{otherwise}.
\end{array}
\right.
\end{equation}
Thus
\beq\label{distributionfunction}
f_0(x,y) = \left\{ \begin{array}{ll}
\displaystyle\frac{1}{h} \sqrt{\frac{\pi}{4 D \tstep}}\, 
 \mathrm{erfc}\left(\frac{x}{\sqrt{4 D \tstep}} \right) 
\left(1-\pdhfrac{|y|}{h}\right), & 
\mbox{for} \; -h < y < h,\\[3mm]
0, & \mbox{otherwise}.
\end{array}
\right.
\eeq
It is important to note that our expansions (\ref{blah5}) assume 
that the probability $p_{i,j}$ of being in compartment 
$\mathcal{C}_{i,j}$ should be a continuous extension of $p(x,y)$ 
evaluated at $(-ih,jh)$ which is in the center of the right side 
of the compartments. If the expansions (\ref{blah5}) were  
taken in the center of the compartments (evaluated at $(-ih-h/2,jh)$) 
the result would be that $C_1 = -\Lambda/2$ and thus a shift in 
the continuous expected probability density curve over the 
interface is seen resulting in an apparent error of 
$h\bar{p}_x/2+O(h^2)$ as $\tstep\rightarrow 0$ on the 
interface \cite{Flegg:2012:TRM}. 
This error can therefore be reduced by refining the compartments near 
the interface (remembering that this might mean also reducing $\tstep$ 
such that $D\tstep \sim h^2$).  

The  derivation above  was conducted under the assumption $D\tstep \sim
h^2$. Let us now assume $D \tstep \ll h^2$ instead. Then
$\Lambda= \pdhfrac{h}{\sqrt{D\tstep}}$ 
and $\Phi$ (given by (\ref{results1d}) as $2 \Lambda/\sqrt{\pi}$) 
are no longer of
order 1, and the above derivation may fail. In particular, we retain terms
of order $h$ while ignoring terms of order $\Phi h^2$. 
If we are interested in very small $\tstep$ the errors associated with
this mismatch between $\tstep$ and $h$ may be reduced by
replacing  $F(y)$ by the step function
\begin{equation}\label{simple2} \bar{F}(y) = \left\{ \begin{array}{ll}
\displaystyle \pdhfrac{1}{h}, & 
\mbox{for} \; -h/2 < y < h/2,\\[3mm]
0, & \mbox{otherwise},
\end{array}
\right.
\end{equation}
which gives $O(h)$ accuracy to (\ref{blah4.5}). This function is also
easier to implement in the case of corners which will be discussed in the
next section. Using (\ref{simple2}) instead of (\ref{simple}), equation
(\ref{distributionfunction}) reads as follows
\beq\label{distributionfunction2}
f_0(x,y) = \left\{ \begin{array}{ll}
\displaystyle\frac{1}{h} \sqrt{\frac{\pi}{4 D \tstep}}\, 
 \mathrm{erfc}\left(\frac{x}{\sqrt{4 D \tstep}} \right), & 
\mbox{for} \; -h/2 < y < h/2,\\[3mm]
0, & \mbox{otherwise}.
\end{array}
\right.
\eeq
These issues will be discussed further in Section \ref{examples}.

\subsection{Interface corners}
\label{seccorners}

\begin{figure}[t]
\centerline{
\raise 4.685cm \hbox{\raise 0.9mm \hbox{(a)}}
\hskip -5.6mm
\epsfig{file=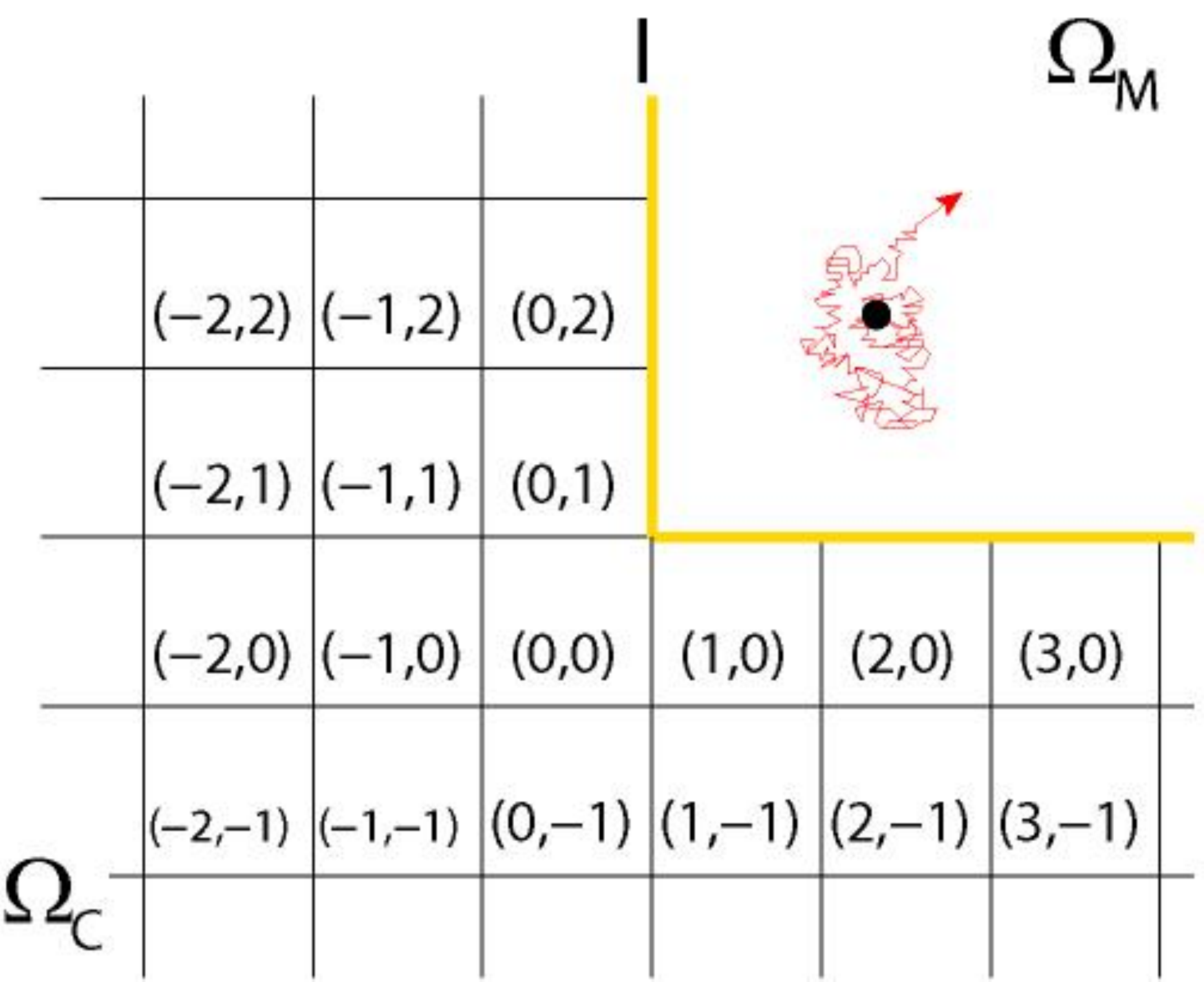,height=4.685cm}
\hskip 1cm
\raise 4.685cm \hbox{\raise 0.9mm \hbox{(b)}}
\epsfig{file=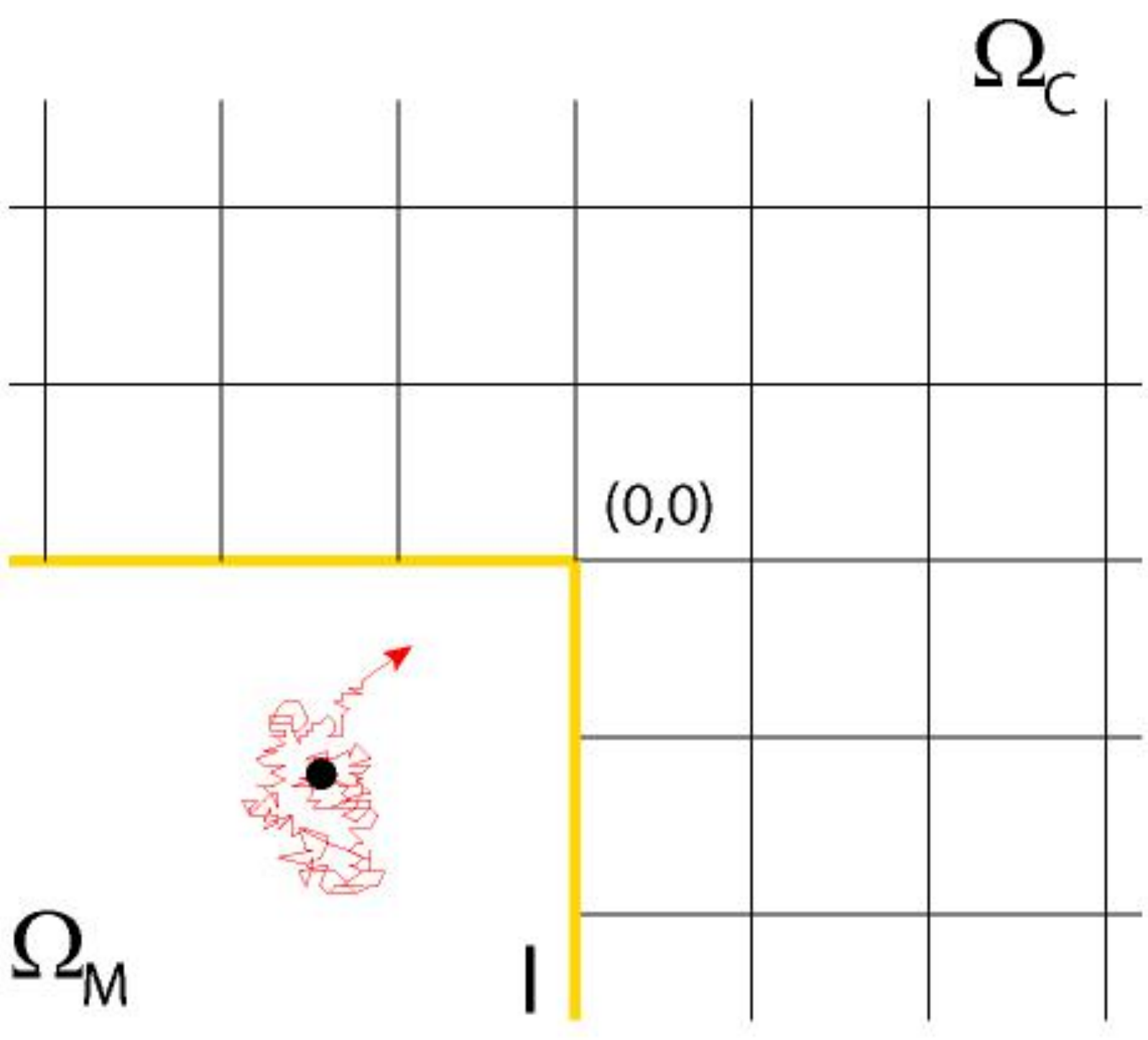,height=4.685cm}
}		
\caption{{\rm (a)} Illustration of corner geometry in two-dimensions.
Compartment indices are denoted as in Section $\ref{seccorners}.$
{\rm (b)} The corner geometry used in illustrative numerical simulations
in Section $\ref{numcorner}.$}
\label{cornerfig}
\end{figure}

The analysis in the previous section does not extend trivially to 
the case when there are corners in the interface $I$. Indeed, when 
the interface is not perfectly flat the previous section is invalid. 
However, since we are restricting our analysis to regular cubic
lattices in $\Omega_C$, the interface must be made up of a 
series of straight edges connected at right angles. Let us consider
the compartments  
$\mathcal{C}_{i,j}$ assigned to the region 
$((i-1)h,ih) \times ((j-1)h,jh)$, where the indices 
$(i,j) \in \mathbb{Z}\times\mathbb{Z} \setminus
\mathbb{N}\times\mathbb{N}$ (i.e. they fill the complement of 
the positive quadrant). The considered geometry is represented 
graphically in Figure \ref{cornerfig}(a). 

Molecules cannot leave compartment $\mathcal{C}_{0,1}$ using (\ref{simple}). 
This is because there is a non-zero probability 
that the molecule crosses diagonally to the region of $\mathcal{C}_{1,0}$ 
and vice versa. This diagonal movement is prohibited by the rules of 
the compartment region but is allowed in the molecular region. Furthermore, 
molecules in $\Omega_M$ may move into $\mathcal{C}_{0,0}$ during one 
time step. Typically, molecules in compartments will not be able 
to move diagonally out of $\mathcal{C}_{0,0}$. Whilst it may be 
possible to make diagonal motion an exception for $\mathcal{C}_{0,0}$, 
atypical functions placing molecules into $\Omega_M$ from 
$\mathcal{C}_{0,0}$, $\mathcal{C}_{0,1}$ and $\mathcal{C}_{1,0}$ must 
be determined and also rules for how molecules in $\Omega_M$ close to the 
corner 
migrate diagonally, left or down into $\Omega_C$ must be determined. It 
is important 
to note that the added complexity to these corner compartments is not 
trivial like the case when the compartments form a straight interface. 
If we attempted to generalize (\ref{simple}), then
complex distributions would have to be sampled from to place molecules from 
$\mathcal{C}_{0,1}$ and $\mathcal{C}_{1,0}$ into $\Omega_M$ and several
tests would have to be performed on molecules in $\Omega_M$ (to see if 
they meet criteria for diagonal migration into $\Omega_C$, or to determine if 
they move down or left over the interface). Given this complication, 
we find it reasonable to use (\ref{simple2})
instead of (\ref{simple}) which means that we use (\ref{distributionfunction2})
instead of (\ref{distributionfunction}). Distribution 
(\ref{distributionfunction2})
does not allow for molecules to leak between 
$\mathcal{C}_{0,1}$ and $\mathcal{C}_{1,0}$ but at the cost 
of accuracy in the local region around the corner. 

The treatment of molecules at the corner is therefore described by the 
following rules. Molecules in compartments $\mathcal{C}_{0,1}$ and 
$\mathcal{C}_{1,0}$ have the propensities given by (\ref{qbar}), namely
$$
\Phi \frac{D}{h^2} \, \mathcal{N}_{0,1},
\qquad
\Phi \frac{D}{h^2} \, \mathcal{N}_{1,0},
$$
where $\mathcal{N}_{0,1}$ (resp. $\mathcal{N}_{1,0}$) is
the number of molecules in the compartment $\mathcal{C}_{0,1}$ 
(resp. $\mathcal{C}_{1,0}$). Molecules from $\mathcal{C}_{0,0}$ 
may not enter $\Omega_M$ directly.
Molecules from $\Omega_M$ that land in $\mathcal{C}_{0,0}$ 
or satisfy condition (b) in Section \ref{secomc}
for both parts $\{x = 0\}$ and $\{y=0\}$
of the interface $I$ are placed at random (with probability
1/2) in $\mathcal{C}_{1,0}$ or $\mathcal{C}_{0,1}$. 
Molecules in the compartment $\mathcal{C}_{0,1}$ 
migrating to $\Omega_M$ are placed according 
to the distribution (\ref{distributionfunction2}).
Molecules in the compartment $\mathcal{C}_{1,0}$ migrating to 
$\Omega_M$ are placed according to the distribution 
\beq\label{distributionfunction3}
f_0(x,y) = \left\{ \begin{array}{ll}
\displaystyle\frac{1}{h} \sqrt{\frac{\pi}{4 D \tstep}}\, 
 \mathrm{erfc}\left(\frac{y}{\sqrt{4 D \tstep}} \right), & 
\mbox{for} \; -h/2 < x < h/2,\\[3mm]
0 & \mbox{otherwise},
\end{array}
\right.
\eeq
which can be obtained from the distribution (\ref{distributionfunction2}) 
by exchanging the variables $x$ and $y$. Thus, in both cases,
we use (\ref{trans1d}) perpendicular from their respective 
interfaces and the step distribution (\ref{simple2}) tangentially 
along each respective interface. 

Since $D \tstep < h^2$ molecules that are in compartments
$\mathcal{C}_{i,0}$ and $\mathcal{C}_{0,j}$ ($i,j\geq 2$) are not
significantly affected by the corner within one time step. We
therefore use the probability distribution
(\ref{distributionfunction2}) for these compartments. 
Thus we use the distribution (\ref{simple2}) for molecules leaving 
the corner compartments, and the triangle distribution 
(\ref{simple}) for molecules leaving all other boundary compartments.

Finally, we note that we also tested the alternative of using a one-sided 
triangle distribution for the corner compartments
$\mathcal{C}_{1,0}$ or $\mathcal{C}_{0,1}$. It produced results which were 
indistinguishable numerically from the distribution (\ref{simple2}), 
although this one-sided distribution includes a small bias of particles 
away from the corner.

\subsection{Parameters for a flat interface in $N$ dimensions} 
\label{Ndim}
The derivation presented in Section \ref{flat2d} can be used for
flat interfaces in arbitrary dimensions. One can show that the parameters 
$\Phi_j = \Phi$ and $\Psi$ are independent of the number 
of dimensions $N$ of the domain for a flat interface with a regular 
cubic compartment arrangement in $\Omega_C$ (and therefore take 
the values derived in Section \ref{flat2d}). Furthermore one can 
show that the distribution $f_0(\mathbf{x})$ for placing molecules 
in $\Omega_M$ is the product of $N$ distributions separating the 
coordinates 
\begin{equation}
f_0(x_1,x_2,\ldots, x_N) = F^{\perp}(x_1)\prod_{i=2}^{N}F^{\parallel}(x_i). 
\label{NDresult}
\end{equation}
Here, the distribution $F^{\perp}(x_1)$ is
for the coordinate $x_1$ perpendicular to the interface 
$I = \{x_1 = 0\}$. It is given (see equations (\ref{trans1d})
and (\ref{distributionfunction}))
by 
\begin{equation}
 F^{\perp}(x_1) = \sqrt{\frac{\pi}{4D\tstep}}\mathrm{erfc}
 \left( \frac{x_1}{\sqrt{4D\tstep}}\right).
\end{equation}
The remaining $N-1$ identical 
distributions for each coordinate $x_i$, 
$i=2,\ldots,N$, tangential to the interface $I$, are denoted 
as $F^{\parallel}(x_i)$ in equation (\ref{NDresult}).
If the origin is placed at the center of compartment in question,
then one can follow the derivation presented in Section \ref{flat2d}
to obtain $F^{\parallel}(x_i) = F(x_i)$ given by (\ref{simple}).
Equation (\ref{NDresult}) indicates that tangential 
coordinates should be chosen independently from each other.

\section{Illustrative numerical examples}
\label{examples}

In this section we present simple diffusion simulations using the TRM 
to demonstrate its accuracy and convergence. We will compare results
computed by distributions (\ref{distributionfunction}) and
(\ref{distributionfunction2}).

\subsection{Straight interface} 
\label{secstraight}
We will consider a diffusing molecule which starts at the origin 
at (dimensionless) time $t=0$ and diffuse with (dimensionless) 
diffusion constant $D=1$ in the semi-infinite two-dimensional 
domain $\Omega = (0,\infty)\times(0,\infty)$. Boundary 
$\partial \Omega$ will be considered reflective. 

The probability distribution, $P(x,y,t)$, to find 
the molecule at time $t$ given its initial position at the origin 
evolves according to the partial differential equation (PDE)
\begin{equation}\label{PDEsimple}
 \frac{\partial P}{\partial t} 
 = 
 \frac{\partial^2 P}{\partial x^2}
 +
 \frac{\partial^2 P}{\partial y^2},
\end{equation}
with the initial condition $P(x,y,0) = \delta(x,y)$ where
$\delta$ is the Dirac delta function. Using no-flux 
(reflective) boundary condition ($\nabla P\cdot \mathbf{\hat{n}} = 0$), 
equation 
(\ref{PDEsimple}) can be solved as
\begin{equation}\label{PDEsol}
 P(x,y,t) = \frac{1}{\pi t}\exp\left( \frac{-(x^2+y^2)}{4t}\right).
\end{equation}
We shall simulate this diffusion process stochastically using the TRM
where $\Omega$ is divided into 
\begin{equation}
\Omega_M = (0,0.5)\times(0,\infty)
\qquad 
\mbox{and}
\qquad
\Omega_C = (0.5,\infty)\times(0,\infty).
\label{strdiv} 
\end{equation}
We run $N_0 = 2\times10^5$
realizations of the TRM for the molecule starting at the origin and 
note its state (either in the molecular regime or compartment regime) 
at each time step until $t=1$. We simulate the compartment regime using 
a square lattice with non-dimensional compartment spacings $h=0.05$, 
$h=0.1$ and $h=0.25$. Since our analysis uses the assumption $D \tstep \sim h^2$, we use the time steps $\tstep = 0.0004$, $\tstep = 0.0016$ and $\tstep = 0.01$ respectively with $D=1$ such that $h/\sqrt{D \tstep} = 2.5 \sim O(1)$.

\begin{figure}[t]
	\centering
	\scalebox{0.43}{
		\includegraphics{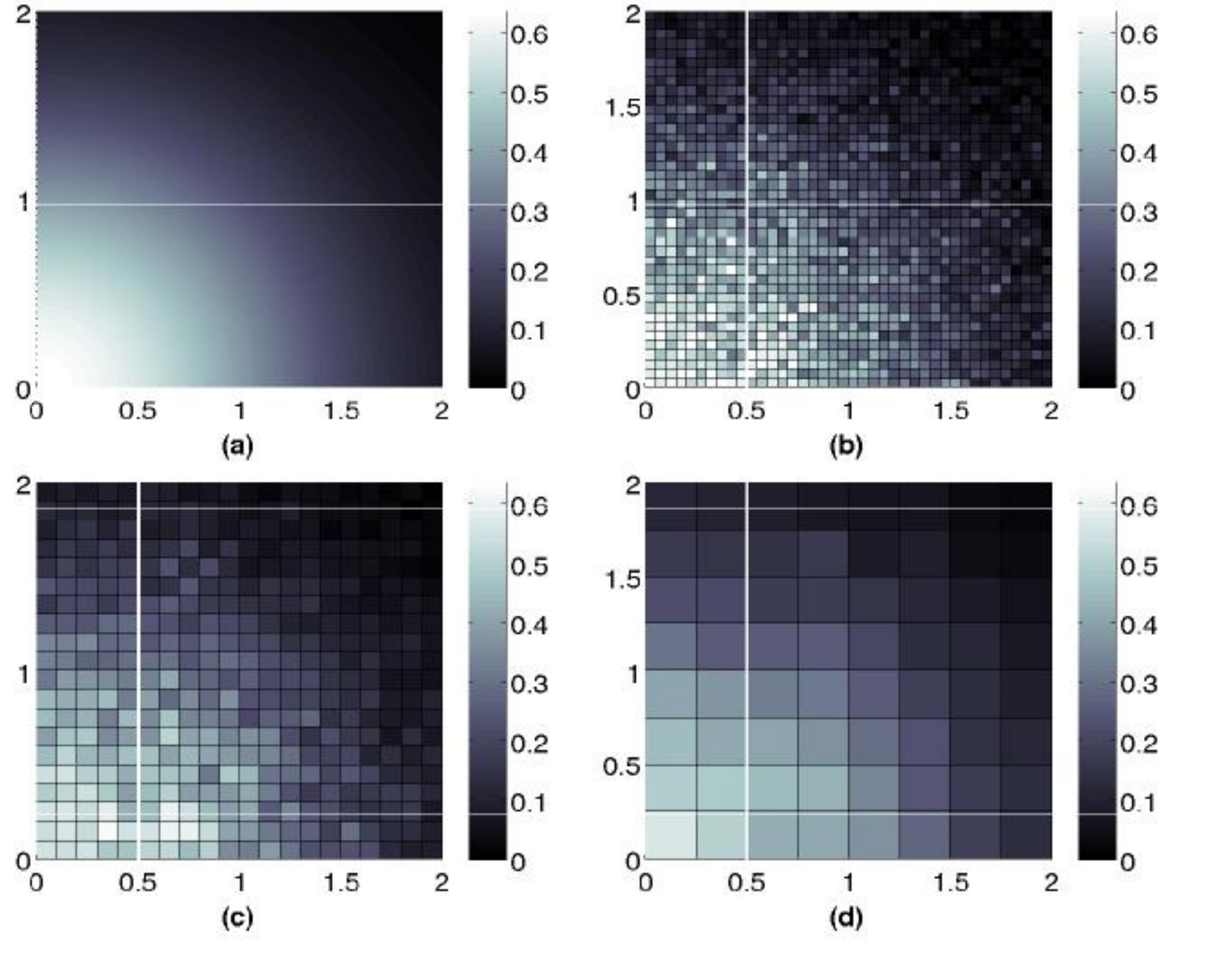}}
	\caption{Probability distribution at time $t=0.5$ 
	estimated using 
	$N_0 = 2\times10^5$ realisations of the TRM method 
	for the domain partition $(\ref{strdiv})$.
	{\rm (a)} The expected distribution found by $(\ref{PDEsol})$. 
	{\rm (b)} TRM simulation with compartment size $h=0.05$ and $\Delta t = 0.0004$. 
	{\rm (c)} TRM simulation with compartment size $h=0.1$ and $\Delta t = 0.0016$. 
	{\rm (d)} TRM simulation with compartment size $h=0.25$ and $\Delta t = 0.01$. 
	$\Omega_C$ can be seen in (b)-(d) on the right and 
	$\Omega_M$ to the left of the white solid line. 
	These simulations were done using sampling 
	$(\ref{distributionfunction})$.
	}\label{dist_straight}
\end{figure}

At $t=0.5$ the molecule positions (for each realisation) are binned 
according to their compartment (or in the case of the molecular regime, 
counted in bins of area $h^2$) and a plot of these bin copy 
numbers divided by $N_0h^2$ is produced to show the approximate 
probability density that is generated by the TRM. These probability densities 
are shown for comparison against the exact solution (\ref{PDEsol}) 
in Figure \ref{dist_straight} for each compartment size $h$. 
The distributions generated using the TRM match well with the 
expected distribution and appear to be more accurate as $h$ is decreased.

To better visualize the accuracy of the TRM, we define 
the error function
\begin{equation}
\mbox{Error}(t) 
=
\frac{C_{TRM}(t)}{N_0} 
-
\iint_{\Omega_C} P(x,y,t) \ \mathrm{d}x \ \mathrm{d}y,
\label{measurederror}
\end{equation}
where $C_{TRM}(t)$ is the number of realisations of the TRM
which have the molecule positioned in $\Omega_C$ at time $t$.
Therefore, the fraction $C_{TRM}(t)/N_0$ is the approximation of
$\iint_{\Omega_C} P(x,y,t) \ \mathrm{d}x \ \mathrm{d}y$
and the error (\ref{measurederror}) measures the accuracy
of this approximation.

\begin{figure}[t]
	\centering
	\includegraphics[width=14.2cm,height=6.1cm]{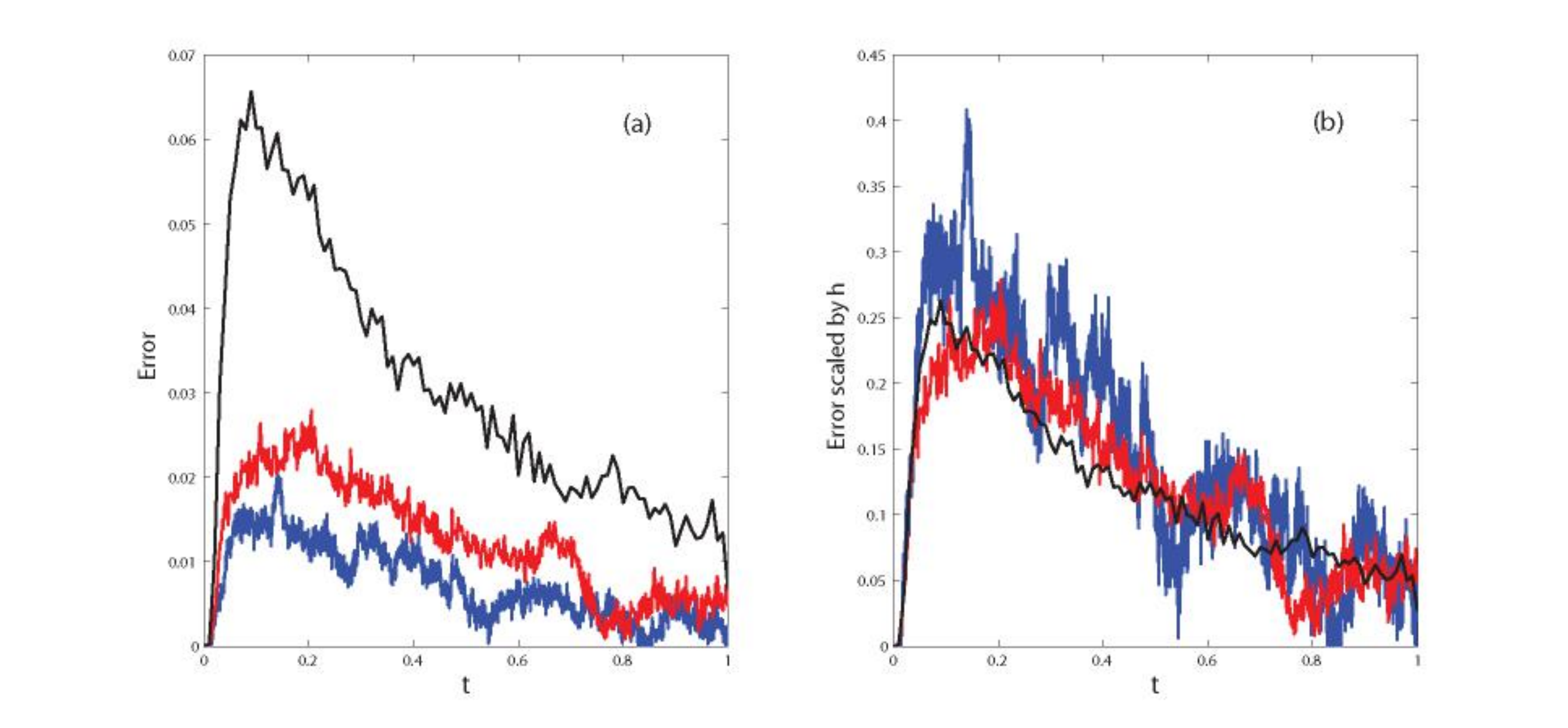}
	\caption{{\rm (a)}
	The error of the TRM defined by $(\ref{measurederror})$
	for $h=0.05$ and $\Delta t = 0.0004$ (blue line), $h=0.1$  and $\Delta t = 0.0016$ (red line) and 
	 $h=0.25$  and $\Delta t = 0.01$ (black line). TRM simulations are taken with
	$\Omega_C = (0.5,\infty)\times(0,\infty)$
        and $\Omega_M = (0,0.5)\times(0,\infty)$.
	  These simulations were done using sampling
	   $(\ref{distributionfunction})$.\quad
	   {\rm (b)} Results from panel (a) scaled by $h$.
	  }
   \label{compartmenterror}
\end{figure}%

In Figure \ref{compartmenterror}(a) we present the error 
(\ref{measurederror}) as a function 
of time for the $h=0.05$, $h=0.1$ and $h=0.25$ simulations. There 
is a maximum in this error around $t \sim 0.1$ for all simulations. 
The predicted error on the boundary (see Section \ref{flat2d}) 
is proportional to the net flux over the interface ($h\bar{p}_x/2+O(h^2)$). 
We note that this flux reaches a maximum (according to the exact
solution (\ref{PDEsol})) at around $t=1/24$. The reason why the maximum 
of our measured error does not match up with this time is because the 
error at the interface is described by a small discontinuity 
in the distribution 
on the boundary, this discontinuity then diffuses into each of the 
subdomains. After $t=1/24$ the discontinuity is reduced, thereby 
reducing this bias effect but there is still some time before 
the distributions are corrected by diffusion. It is for this reason 
that coupling at the interface correctly is so crucial. A small bias 
in the flow over the interface in one direction can lead to an avalanching 
effect on the distribution. Whilst the TRM cannot eliminate the error 
that is associated with changing of regime entirely, it does optimize 
the error. We also expect the error, however, to be proportional 
to $h$. We can see that this is the case by plotting the error 
(\ref{measurederror})
divided by $h$ for each simulation (Figure 
\ref{compartmenterror}(b)). Each of the three curves in Figure 
\ref{compartmenterror}(b) approximately overlay implying that 
the leading order term of the error is $O(h)$. This means that in the continuous 
limit $h\rightarrow 0$ the error that is due to the TRM appoximately converges 
linearly with $h$, as expected.

\begin{figure}[t]
	\centering
	\scalebox{0.43}{
		\includegraphics{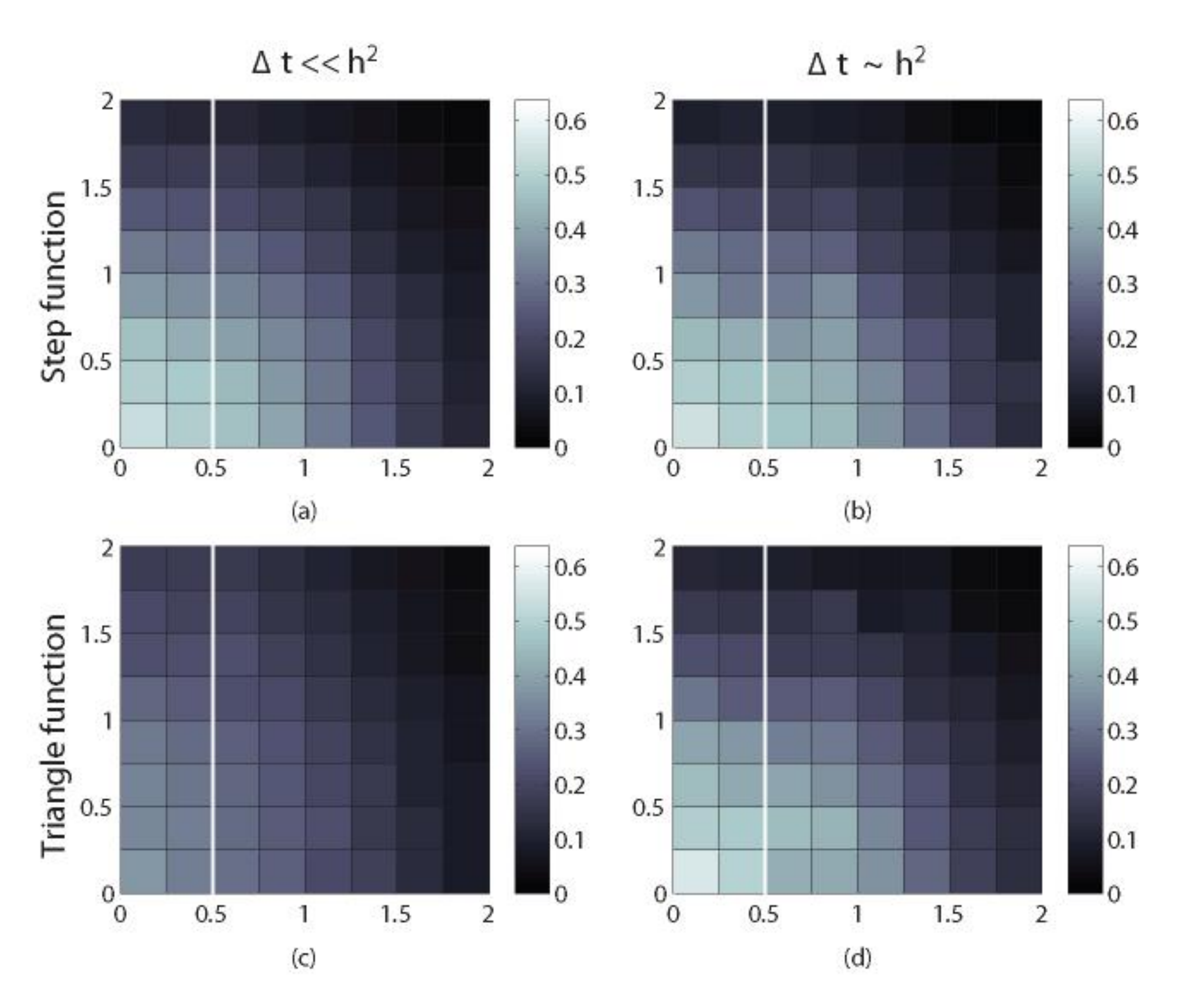}}
	\caption{
	Probability distribution at time $t=0.5$ 
	estimated using 
	$N_0 = 2\times10^5$ realisations of the TRM method 
	for the domain partition $(\ref{strdiv})$.
	In all simulations $h=0.25$ and $D=1$.
	$\Omega_C$ can be seen in (a)-(d) on the right and 
	$\Omega_M$ to the left of the white solid line denoting
	interface $I$. TRM simulations with 
	{\rm (a)} $D \Delta t = 10^{-5} \ll h^2$ 
	and $(\ref{distributionfunction})$;
	{\rm (b)} $D \Delta t = 0.01  \sim h^2$ 
	and $(\ref{distributionfunction})$; \hfill\break
	{\rm (c)} $D \Delta t = 10^{-5} \ll h^2$ 
	and $(\ref{distributionfunction2})$;
	{\rm (d)} $D \Delta t = 0.01  \sim h^2$ 
	and $(\ref{distributionfunction2})$. 
	}\label{tangcompare}
\end{figure}

The purpose of the TRM is to match the concentration and perpendicular 
flux of molecules at the boundary in the most optimal way given 
small $h$ and small $\Delta t$. In our analysis, we restricted
the algorithm parameters
to the case $D \Delta t \sim h^2$. In some applications, this might
not be a preferred parameter regime because $\Delta t$ determines the resolution 
of the microscopic region $\Omega_M$ and if $D \Delta t \sim h^2$
then this resolution is no better than the compartment-based regime. 
One must be careful in this case since for $D \Delta t \ll h^2$, 
the parameter $\Phi = 2 \Lambda/\sqrt{\pi}$ (defined by (\ref{ratiophis})) 
may become significantly
larger than 1. Previously, higher order terms of the form $\Phi h^2$
in (\ref{blah3.3})--(\ref{blah4.2})
(which were ignored as ``too small'') now become dominant over terms that 
are of the order of $h$. This manifests itself into an increase in 
the dispersion inside the boundary layer near the interface 
in the tangential direction. This effect can be seen in Figure 
\ref{tangcompare}(c). 

The increase in the dispersion in the tangential direction may also
be explained by following 
the TRM mechanism. Consider molecules in $\Omega_M$ close to 
the interface between $\Omega_M$ and $\Omega_C$. If these molecules 
are sufficiently close to the interface, they are likely to 
be absorbed into a compartment in $\Omega_C$. 
To compensate for this rapid absorption of molecules, the compartments 
on the interface must return these molecules to the boundary layer 
at comparable rates. Since, as $\Delta t\rightarrow 0$, the boundary 
layer gets thinner, this increases the rate of resorption of molecules 
having just come from $\Omega_C$. If a 
molecule enters an interfacial compartment $\mathcal{C}_{0,j}$ 
near the  boundary to the adjacent compartment $\mathcal{C}_{0,j-1}$ 
then this molecule effectively jumps a distance $h/2$ in the tangential 
direction (because its position in $\Omega_C$
can be considered as the center of $\mathcal{C}_{0,j}$).
This molecule is then rapidly interchanged between $\Omega_M$ 
and $\mathcal{C}_{0,j}$ each time with a new tangential coordinate until 
the tangential coordinate falls out of line with $\mathcal{C}_j$ or 
the molecule diffuses away from the interface. The former of these 
two options occurs more often if $\Delta t$ is small since 
the tangential coordinate is rapidly resampled until the 0.25 chance 
of being sampled outside the compartment $\mathcal{C}_j$ is realized 
if we use (\ref{simple}). In these situations, it is better to use 
(\ref{simple2}) for the tangential coordinate. This is because, 
(\ref{simple2}) restricts the molecule initiation to the compartment
from which it came. Furthermore, 
one can reduce this effect by reducing the size of $h$. Reducing $h$ helps 
to reduce $\Phi$. 

It is possible to show that this numerical artefact 
is improved, if $h$ cannot be reduced, by using step function (\ref{simple2}) 
instead of triangle function (\ref{simple}) to sample molecule positions tangentially 
to the interface. However, if $D \Delta t \sim h^2$ then 
(\ref{simple}) offers the best results. This is demonstrated 
in Figure \ref{tangcompare}. In Figure \ref{tangcompare}, 
the least amount of artificial dispersion is achieved if 
$D \Delta t \sim h^2$ using the triangle function 
sampling (\ref{simple}) but the triangle function 
sampling introduces more severe artificial dispersion 
along the interface than step function sampling (\ref{simple2}) 
if $D \Delta t \ll h^2$.

\subsection{Interface with a corner}
\label{numcorner}
To demonstrate that the TRM produces good results when there 
is a corner in the interface, we present also the results 
of TRM simulations of the same problem as in Section
\ref{secstraight} with subdomains redefined as follows
\begin{equation}
\Omega_M = (0,0.5) \times (0,0.5), 
\qquad
\Omega_C = (0,0.5) \times (0.5, \infty)
\cup (0.5,\infty) \times (0, \infty).
\label{cornerpartition}
\end{equation}
The details of the TRM implementation of corners are discussed in Section
\ref{seccorners}. The corner is oriented as in Figure \ref{cornerfig}(b),
i.e. the TRM implementation presented in Section \ref{seccorners} for the
corner orientation in Figure \ref{cornerfig}(a) is adjusted (by 
a simple rotation) to the corner orientation in Figure \ref{cornerfig}(b).

\begin{figure}[t]
	\centering
	\scalebox{0.43}{
		\includegraphics{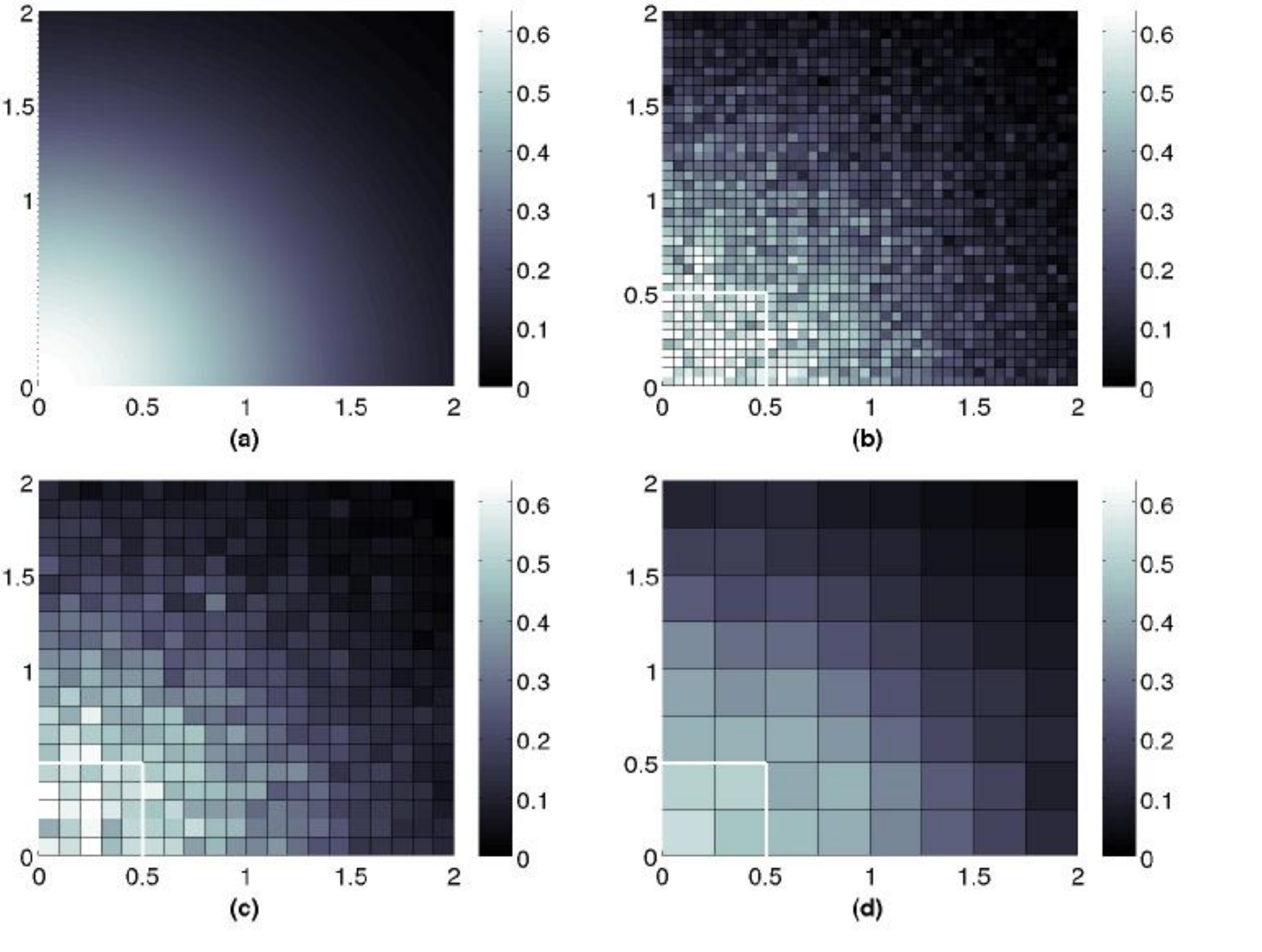}}
	\caption{Probability distribution at time $t=0.5$ 
	estimated using 
	$N_0 = 2\times10^5$ realisations of the TRM method 
	for the domain partition $(\ref{cornerpartition})$.
	{\rm (a)} The expected distribution found by the exact
	solution $(\ref{PDEsol})$. 
	{\rm (b)} TRM simulation with compartment size $h=0.05$ and $\Delta t = 0.0004$. 
	{\rm (c)} TRM simulation with compartment size $h=0.1$ and $\Delta t = 0.0016$. 
	{\rm (d)} TRM simulation with compartment size $h=0.25$ and $\Delta t = 0.01$.
	$\Omega_C$ can be seen in (b)-(d) outside of $\Omega_M$ 
	which is boxed by the white solid line (interface $I$) 
	about the origin. 
	These simulations were done using sampling 
	$(\ref{distributionfunction2})$--$(\ref{distributionfunction3})$ near the corners
	according to the approach described in Section $\ref{seccorners}$.
	}\label{dist_corner}
\end{figure}

\begin{figure}[t]
	\centering
	\includegraphics[width=14cm,height=6cm]{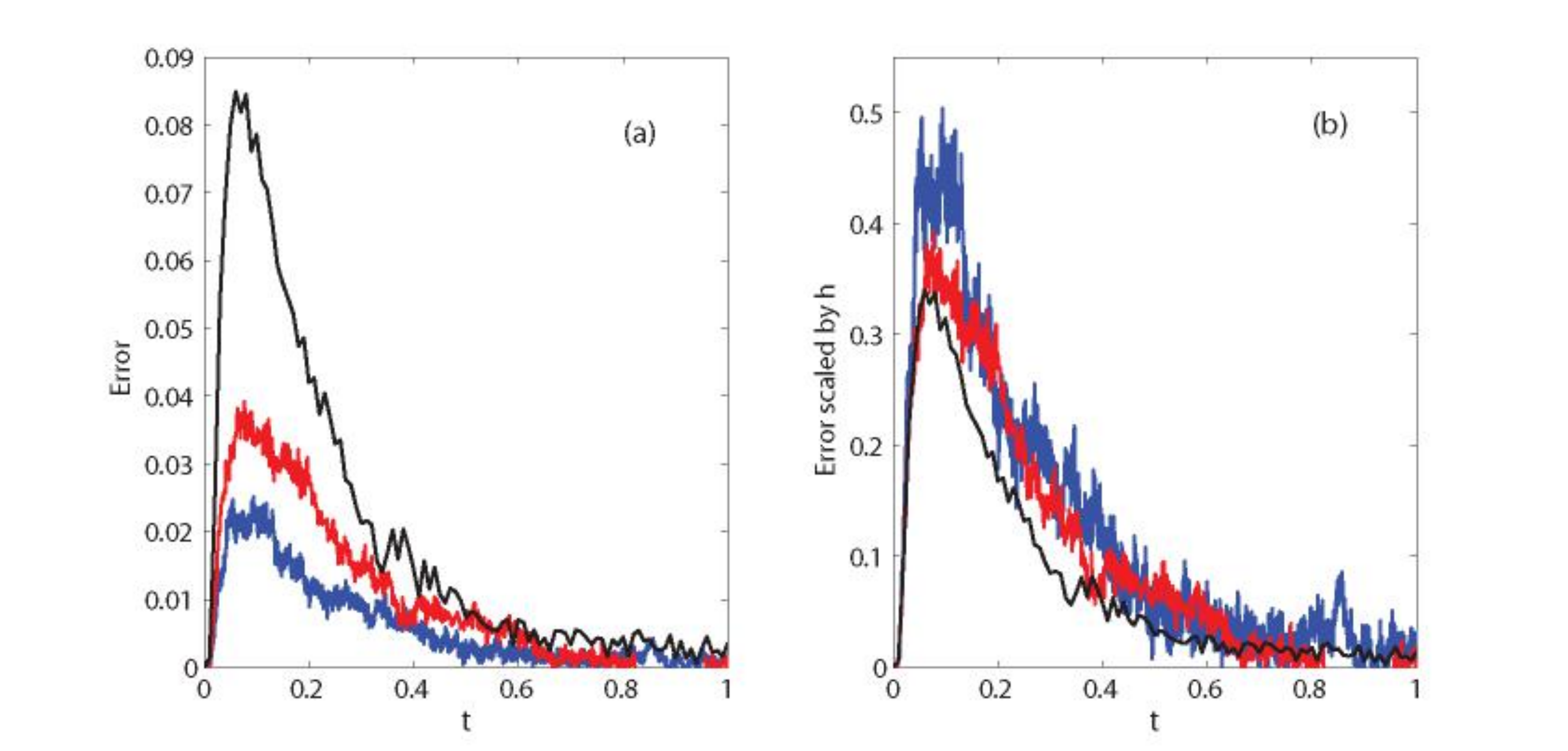}
	\caption{{\rm (a)}
	The error of the TRM defined by $(\ref{measurederror})$
	for $h=0.05$ and $\Delta t = 0.0004$ (blue line), $h=0.1$ and $\Delta t = 0.0016$ (red line) and 
	 $h=0.25$ and $\Delta t = 0.01$ (black line). TRM simulations are taken 
	with $\Omega_M$ and $\Omega_C$ defined in
	$(\ref{cornerpartition})$. 
	{\rm (b)} Results from (a) scaled by $h$.
	}\label{compartmenterrorcorner}
\end{figure}

In Figure \ref{dist_corner} we present the distributions 
that result using the TRM for the domain partition
(\ref{cornerpartition}). These distributions are calculated in 
the same way as those distributions found in Figure \ref{dist_straight}. 
The distributions are plotted, similarly, at $t=0.5$. There is 
good agreement with the expected distribution (Figure \ref{dist_corner}(a))
especially for small $h$. 
In Figure \ref{compartmenterrorcorner}(a), we present the error 
(\ref{measurederror}) as a function of time for the $h=0.05$, 
$h=0.1$ and $h=0.25$ simulations.
To verify that the error due to the TRM is still $O(h)$ when 
the interface has a corner in it, Figure \ref{compartmenterrorcorner}(b) 
shows the error for the TRM simulations scaled by $h$.

\begin{figure}[t]
	\centering
	\includegraphics[width=14cm,height=6cm]{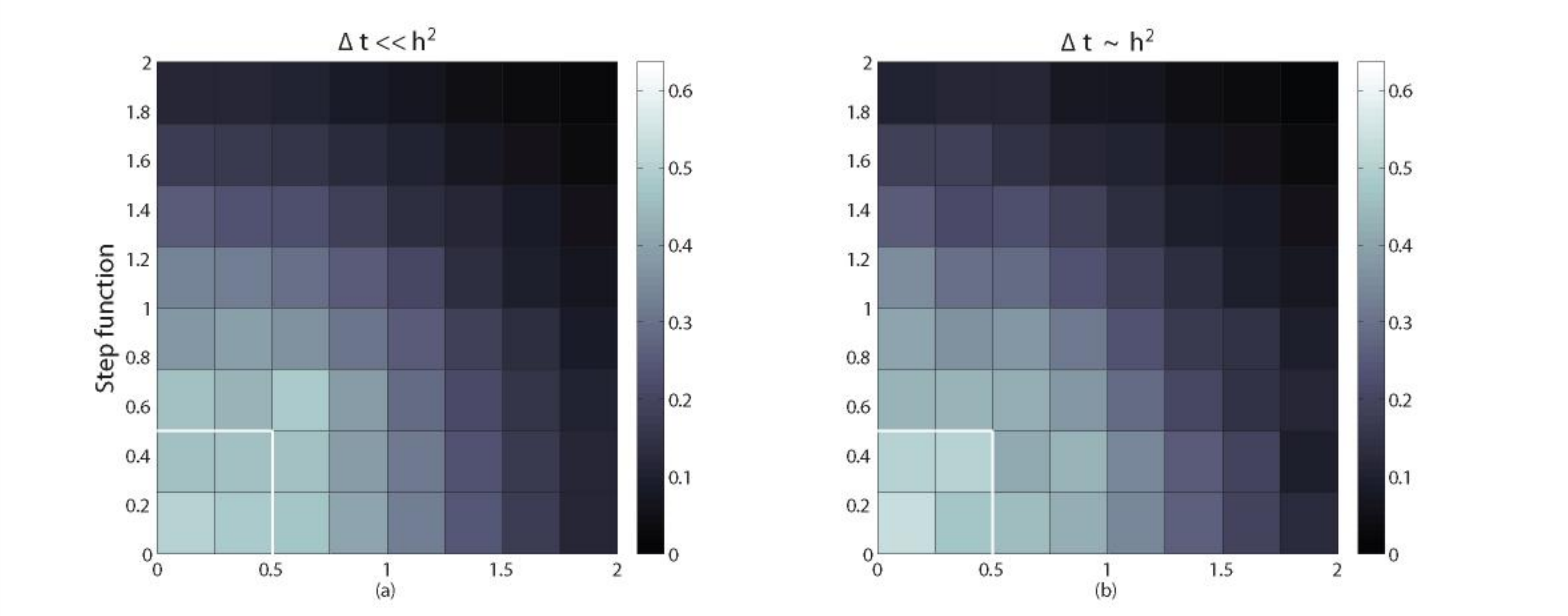}
	\caption{Probability distribution at time $t=0.5$ 
	estimated using 
	$N_0 = 2\times10^5$ realisations of the TRM method 
	for the domain partition $(\ref{cornerpartition})$.
	In both simulations $h=0.25$ and $D=1$. 
	$\Omega_C$ can be seen outside of $\Omega_M$ 
	which is boxed by the white solid line denoting interface
	$I$. TRM simulations with
	{\rm (a)} $D \Delta t = 10^{-5} \ll h^2$; 
	{\rm (b)} $D \Delta t = 0.01  \sim h^2$. 
        These simulations were done using sampling 
	$(\ref{distributionfunction2})$--$(\ref{distributionfunction3})$ near the corners
	according to the approach described in Section $\ref{seccorners}$.
	}\label{tangcomparecorner}
\end{figure}

In Figure \ref{tangcomparecorner}(a), we again see the artefact for small $\tstep$ in the compartment described by the region 
$x\in [0.5,0.75)$, $y\in [0.5,0.75)$; 
an unexpectedly large number of molecules. This is because diffusion 
is biased tangentially along both sides of the interface (due 
to the large $h$ and small $\Delta t$ as we discussed in
Section \ref{secstraight}) and these molecules gather 
at the corner.

\section{Discussion}
\label{secdiscussion}
In this paper we presented an analysis of the TRM 
on regular lattices in dimensions larger than one. We have derived 
TRM parameters to 
simulate diffusing molecules migrating over interface $I$
that separates domain $\Omega_C$ in which molecules are constrained 
to a lattice and domain $\Omega_M$ whereby molecules can diffuse 
in continuous space. The TRM algorithm is presented in
Table \ref{TRMalg}. We considered two cases 
$D \Delta t \ll h^2$ and $D \Delta t \sim h^2$.
We showed that the parameter $\Phi_{i,j}$ defined in (\ref{qbar})
can be chosen as in the one-dimensional case (\ref{results1d}).
The distribution $f(\mathbf{x})$ for 
placing molecules in $\Omega_M$ was presented in equation
(\ref{NDresult}). In particular, each tangential direction
to the interface may be treated independently and in the
same way. We presented two different approaches (\ref{simple}) 
and (\ref{simple2}) for sampling tangential directions. 

In the case $D \Delta t \ll h^2$, the step function approximation
(\ref{simple2}) is the most appropriate choice. In particular, 
the distribution for placing molecules in $\Omega_M$ is given by 
(\ref{distributionfunction2}) for two-dimensional problems. 
In the case $D \Delta t \sim h^2$, the triangle function approximation
(\ref{simple}) is the most appropriate choice, i.e. the distribution
for placing molecules in $\Omega_M$ is given by 
(\ref{distributionfunction}). Moreover, Figure \ref{tangcompare}
demonstrates that the overall best results (in terms of accuracy)
can be obtained in the case $D \Delta t \sim h^2$ and (\ref{simple}). 
However, imposing the condition $D \Delta t \sim h^2$ over the
whole domain $\Omega_C$ might lead to computationally intensive
simulations. A natural solution to this problem would be to 
use unstructured meshes \cite{Engblom:2009:SSR} where compartments
can be of different sizes. Then we could impose the condition 
$D \Delta t \sim h^2$ for compartments
close the interface $I$ (to maximise accuracy) and 
the condition $D \Delta t \ll h^2$ for compartments
further from the interface (to maximise efficiency).

The last decade has seen a number of different algorithms appear 
in the scientific literature with the purpose of coupling two 
domains in space with different modelling techniques for 
reaction-diffusion processes similar to that of the TRM. Some 
of these algorithms aim to couple mesoscopic stochastic 
simulations on a lattice (compartment-based model) with 
a deterministic PDE-based mean-field description 
\cite{Alexander:2002:ARS,Flekkoy:2001:CPF,Wagner:2004:HCF,Moro:2004:HMS}. 
Most of these algorithms 
include the use of an overlap region and use this overlap 
region to calculate the flux of molecules (and other conserved 
physical quantities) that are flowing between the two regimes. 
This flux is calculated by particle counting and matched to 
the flux condition which forms the boundary condition of 
the deterministic macroscopic approach 
\cite{Alexander:2002:ARS, Flekkoy:2001:CPF,Wagner:2004:HCF, Moro:2004:HMS}. 
New particles are created either in a new compartment 
corresponding to a lattice-point used in the PDE solver 
that is implemented on the deterministic side of the 
interface (``handshaking region'') \cite{Alexander:2002:ARS} 
or in the closest compartment to the position at which 
the deterministic region terminates \cite{Flekkoy:2001:CPF,Wagner:2004:HCF}. 
Whilst these approaches often have a region of overlap 
where the description of particles is transitioning from 
deterministic to compartment-based, these regions are 
typically thin and transition between the descriptions 
is clearly defined in space. Ferm {\it et al.} \cite{Ferm:2010:AAS}
presented a new technique for connecting deterministic 
regions with mesoscopic regions using a smoother coupling 
technique. Their technique involves interfacing a region 
of deterministic description with a stochastic description 
modelled using a tau leaping algorithm. The tau leaping algorithm 
is a quick compartment-based algorithm that does not account 
for all events as they occur but rather updates the system 
at discrete moments in time $\tau$ \cite{Gillespie:2001:AAS}. 
This stochastic simulation 
technique is fast but is only accurate if the concentration 
of molecules is large enough that individual events on smaller 
time scales produce small perturbations that do not affect 
the simulation significantly. This tau leaping algorithm is 
then interfaced with a more accurate compartment-based algorithm 
as the concentration drops.

Geyer {\it et al.} \cite{Geyer:2004:IBD, Gorba:2004:BDS}
developed a method for coupling a deterministic numerical solution 
to a reaction-diffusion PDE with a Brownian dynamics molecular-based 
algorithm. In a similar way to the previous algorithms which couple 
deterministic PDEs and compartment-based algorithms the flux 
is determined at the interface. Molecules are initiated in 
the molecular-based algorithm in one-dimension using the same 
distribution that is stated in this manuscript (\ref{trans1d}) 
(see (11) in \cite{Geyer:2004:IBD}). Indeed this boundary 
condition is generated by considering a ``compartment'' with 
a particular expected number of molecules. Absorption 
of molecules at the interface is done without consideration 
of condition (b) in Section \ref{secomc}. Since this condition 
is necessary in the TRM to generate the correct expected 
flux on both sides, the matching of the flux is instead 
forced manually by the algorithm. Unlike the algorithm 
presented in \cite{Geyer:2004:IBD} the TRM couples two 
stochastic simulations. In the case of the TRM, the expected 
flux cannot be calculated without averaging over time. 
We present the TRM specifically with rules that govern each 
molecule as they cross the interface rather than try to 
introduce conditions that depend on the simulation itself. 
This is a more natural methodology since it should not be 
the case that an individual molecule is influenced by the 
net flux but rather be treated independently from each other.

Franz {\it et al.} \cite{Franz:2012:MRA} recently developed 
a technique for coupling deterministic PDEs with microscopic 
molecular-based simulations. Whilst this technique is different 
from the TRM as it interfaces a deterministic PDE-based
description with a stochastic one, it deserves special 
mention because unlike other techniques that couple 
deterministic systems with stochastic simulations it does 
not use an averaging technique to determine the flux over 
the interface. Rather, it treats each particle as an individual 
as it crosses from the molecular-based simulation and is 
added to the deterministic description in a probabilistic 
sense, where the probability for finding the particle is 
known from the simulation. Transversely, molecules may 
migrate back when they are sampled from the probability 
distribution which is proportional to the continuous 
distribution of particles.

An algorithm similar to the TRM was recently published 
by Klann {\it et al.} \cite{Klann:2012:HSG}. 
In this algorithm, compartment-based and molecular-based 
regions are coupled. Klann {\it et al.} \cite{Klann:2012:HSG}
present no comparison with the TRM and it is not clear how exactly
they couple different regimes. However, several 
differences between their approach and the TRM can be 
identified. 

The analysis presented in \cite{Flegg:2012:TRM} 
and in Section \ref{2Dresults} reveals that some steps 
of the Klann {\it et al.} hybrid method \cite{Klann:2012:HSG} 
are not optimal. For example, in \cite{Klann:2012:HSG},
a molecule migrates from the compartment-based algorithm with 
a propensity which is natural for the lattice. This means 
that they have taken a value of $\Phi = 1$ instead
of $\Phi$ given by (\ref{results1d}). Instead of 
placing the molecule with a distribution given 
by (\ref{distributionfunction}) into the the molecular-based 
domain they place the molecule within the compartment 
that corresponds with a natural extension of the 
compartment-based approach. Furthermore, molecules are 
transferred back into compartment-based molecules without the 
condition (b) in Section \ref{secomc} (this corresponds 
to a value of $\Psi = 1/2$ instead of
$\Psi$ given by (\ref{results1d})). It is our 
experience that somewhat ad hoc or heuristic coupling of this type 
between compartment-based and molecular-based regions, 
despite being simpler to implement, may lead to heavy 
biasing of molecules especially in the case when the 
expected net flux over the interface is high. 
This biasing also depends on the relationship between 
$\tstep$ and $h$.

\bigskip
\bigskip

\noindent
{\bf Acknowledgements.} The research leading to these results has received 
funding from the {\it European Research Council} under the 
{\it European Community's} Seventh Framework Programme 
({\it FP7/2007-2013})/ ERC grant agreement No. 239870.
This publication was based on work supported in part by Award 
No KUK-C1-013-04, made by King Abdullah University of Science 
and Technology (KAUST). Radek Erban would also like to thank 
the Royal Society for a University Research Fellowship;
Brasenose College, University of Oxford, for a Nicholas Kurti 
Junior Fellowship; and the Leverhulme Trust for a Philip 
Leverhulme Prize.

\end{document}